\documentclass[letterpaper,twocolumn,10pt]{article}
\usepackage{usenix2019,epsfig,endnotes}
\usepackage{xcolor,url}

\begin{document}

\date{}

\title{\Large \bf Understanding and taming SSD read performance variability: HDFS case study}

\author{
{\rm Mar\'ia F.\ Borge}\\
University of Sydney
\and
{\rm Florin Dinu}\\
University of Sydney
\and
{\rm Willy Zwaenepoel}\\
EPFL and University of Sydney
} 

\maketitle

\thispagestyle{empty}


\subsection*{Abstract}

In this paper we analyze the influence that lower layers (file system, OS, SSD)
have on HDFS' ability to extract maximum performance from SSDs on the read
path. We uncover and analyze three surprising performance slowdowns induced by
lower layers that result in HDFS read throughput loss.  First, intrinsic
slowdown affects reads from every new file system extent for a variable amount
of time. Second, temporal slowdown appears temporarily and periodically and is
workload-agnostic. Third, in permanent slowdown, some files can
individually and permanently become slower after a period of time.

We analyze the impact of these slowdowns on HDFS and show significant throughput
loss. Individually, each of the slowdowns can cause a read throughput loss of
10-15\%. However, their effect is cumulative. When all slowdowns happen
concurrently, read throughput drops by as much as 30\%. We further analyze
mitigation techniques and show that two of the three slowdowns could be
addressed via increased IO request parallelism in the lower layers.
Unfortunately, HDFS cannot automatically adapt to use such additional
parallelism. Our results point to a need for adaptability in storage stacks.
The reason is that an access pattern that maximizes performance in the common
case is not necessarily the same one that can mask performance fluctuations.

\section{Introduction}

Layering is a popular way to design big data storage
stacks~\cite{shafer2010hadoop}. Distributed file systems
(HDFS~\cite{hdfs-paper}, GFS~\cite{gfs}) are usually layered on top of a local
file system (ext4~\cite{mathur2007new}, Btrfs~\cite{rodeh2013btrfs}, F2FS~\cite{lee2015f2fs}, 
XFS~\cite{sweeney1996scalability}, ZFS~\cite{bonwick2003zettabyte}) running in an OS. This approach is desirable
because it encourages rapid development by reusing existing functionality. An
important goal for such storage stacks is to extract maximum performance from
the underlying storage.  With respect to this goal, layering is a double-edge
sword~\cite{ousterhout2018always}. On one hand, the OS and the file system can compensate for and mask
inefficiencies found in hardware. On the other hand, they can introduce their
own performance bottlenecks and sources of variability~\cite{cao2017performance,mytkowicz2009producing,sites2018benchmarking}.

In this paper we perform a deep dive into the performance of a critical layer
in today's big data storage stacks, namely the Hadoop Distributed File System
(HDFS~\cite{hdfs-paper}). We particularly focus on the influence that lower
layers have on HDFS' ability to extract the maximum throughput that the
underlying storage is capable of providing. We focus on the HDFS read path and
on SSD as a storage media. We use ext4 as the file system due to its
popularity. The HDFS read path is important because it can easily become a
performance bottleneck for an application's input stage which accesses slower
storage media compared to later stages which are often optimized to work fully
in memory~\cite{rdd}.  

Central to our exploration is the storage access pattern of a single HDFS read
request: single threaded, sequential access to large files (hundreds of
megabytes) using buffered IO. This access pattern is simple but tried and
tested and has remained unchanged since the beginnings of HDFS.  It is
increasingly important to understand whether a single HDFS read 
using this access pattern can consistently extract by itself the maximum
throughput that the underlying storage can provide. As many big data processing
systems are heavily IO provisioned and the ratio of cores to disks reaches
1~\cite{nutanix-platforms}, relying on task-level parallelism to generate
enough parallel requests to saturate storage is no longer sufficient. Moreover,
relying on such parallelism is detrimental to application performance since
each of the concurrent HDFS reads is served slower than it would be in
isolation. 

With proper parameter configuration, the HDFS read access pattern is sufficient
to extract maximum performance out of our SSDs.  However, we uncover and
analyze three surprising performance slowdowns that can affect the HDFS read
path at different timescales (short, medium and long) and result in throughput
loss.  All three slowdowns are caused by lower layers (file system, SSDs).
Furthermore, our analysis shows that the three slowdowns can affect not only
HDFS but any application that uses buffered reads. As opposed to related work
that shows worn out SSDs could cause various performance
problems~\cite{mutlu_errors_date12,gunawi2018fail}, our slowdowns occur on SSD
with very low usage throughout their lifetime.

The first slowdown, which we call intrinsic slowdown, affects HDFS reads at
short time scales (seconds). HDFS read throughput drops at the start of every
new ext4 extent read. A variable number of IO requests from the start of each
extent are served with increased latency regardless of the extent size. The
time necessary for throughput to recover is also variable as it depends on the
number of request affected. 

The second slowdown, which we call temporal slowdown, affects HDFS reads at
medium time scales (tens of minutes to hours). Tail latencies inside the drive
increase periodically and temporarily and cause HDFS read throughput loss.
While this slowdown may be confused with write-triggered SSD garbage
collection~\cite{tail_at_scale}, we find, surprisingly, that it appears in a
workload-agnostic manner.  

The third slowdown, which we call permanent slowdown, affects HDFS reads at long
time scales (days to weeks). After a period of time, HDFS read throughput from
a file permanently drops and never recovers for that specific file. Importantly
this is not caused by a single drive-wide malfunction event but rather it is an
issue that affects files individually and at different points in time. The
throughput loss is caused by latency increases inside the drive, but, compared
to temporal slowdown all requests are affected, not just the tail. 

Interestingly, we find that two of the three slowdowns can be completely masked
by increased parallelism in the lower layers, yet HDFS cannot trigger this
increased parallelism and performance suffers. Our results point to a need for
adaptability in storage stacks. An access pattern that maximizes performance in
the common case is not necessarily  the one that can mask unavoidable hardware
performance fluctuations.   

With experiments on 3 SSD based systems, we show that each of the slowdowns we
identified can individually introduce at least 10-15\% HDFS read throughput
degradation. The effect is cumulative. In the worst case, all slowdowns can
overlap leading to a 30\% throughput loss.

The contributions of the paper are as follows:
\begin{itemize}

\item We identify and analyze the intrinsic slowdown affecting reads
at the start of every new ext4 extent for a variable amount of time.

\item We identify and analyze the temporal slowdown that affects reads
temporarily and periodically while being workload-agnostic.

\item We identify and analyze the permanent slowdown affecting reads
from an individual file over time.

\item We analyze the impact of these slowdowns on HDFS performance and show
significant throughput loss.

\item We analyze mitigation techniques and show that two of the three
slowdowns can be addressed via increased parallelism.

\end{itemize}

The rest of the paper continues as follows. \S\ref{sec:background} presents
HDFS architecture and the configuration that we use. \S\ref{sec:methodology}
presents our experimental setup and the metrics used.  \S\ref{sec:IntSlwdown},
\S\ref{sec:TmpSlwdown} and \S\ref{sec:PermSlwdown} introduce and analyze
intrinsic, temporal and permanent slowdown respectively.
\S\ref{sec:discussion} discusses the findings, \S\ref{sec:related} presents
related work and \S\ref{sec:conclusion} concludes.

\section{Background}
\label{sec:background}

\subsection{HDFS Architecture}
HDFS is a user-level distributed file system that runs layered on top of a
local file system (e.g. ext4, zfs, btrfs). An HDFS file is composed of several
blocks, and each block is stored as one separate file in the local file system.
Blocks are large files; a common size is 256 MB.

The HDFS design follows a server-client architecture. The server, called the
NameNode, is a logically centralized metadata manager that also decides file
placement and performs load balancing. The clients, called DataNodes, work at
the level of HDFS blocks and provide block data to requests coming from compute
tasks.   

To perform an HDFS read request, a compute task (e.g. Hadoop mapper) first
contacts the NameNode to find all the DataNodes holding a copy of the desired
data block. The task then chooses a DataNode and asks for the entire block of
data.  The DataNode reads data from the drive and sends it to the task.  The
size DataNode's reads is controlled by the parameter io.file.buffer.size
(default 64KB).  The DataNode sends as much data as allowed by the OS socket
buffer sizes. A DataNode normally uses the sendfile system call to read data
from a drive and send it to the task.  This approach is general and handles
both local tasks (i.e. on the same node as the DataNode) as well as remote
tasks. As an optimization, a task can bypass the DataNode for local reads
(short-circuit reads) and read data directly using standard read system calls.

\subsection{HDFS Access Pattern}
We now summarize the main characteristics of the HDFS read access pattern since
this pattern is central to our work. 

\begin{itemize}

\item {\bf Single-threaded.} One request from a compute task is handled
by a single worker thread in the DataNode. 

\item {\bf Large files.} HDFS blocks are large files (e.g. 128MB, 256MB). Each
HDFS block is a separate file in the local file system.

\item {\bf Sequential access.} The HDFS reads access data sequentially for
performance reasons.

\item {\bf Buffered IO.} HDFS uses buffered IO in the DataNode (via sendfile or
read system calls).
   
\end{itemize}

\subsection{Removing Software Bottlenecks}
\label{sec:decoupling}
We now detail the configuration changes we made to alleviate network and
compute bottlenecks affecting HDFS and to eliminate sources of interference. As
a result, we observed that HDFS can extract the maximum performance that our
SSDs can generate.

\noindent{\bf File system configuration.} We disabled access
time update, directory access time update, and data-ordered journaling (we use
write-back journaling) in ext4. This removes sources of interference so that
we can profile HDFS in isolation.

\noindent{\bf OS configuration.} Small socket buffer sizes
limit the number of packets that the DataNode can send to a task and thus
reduce performance by interrupting disk reads and inducing disk idleness. 
We increase the socket buffer size for both reads and write to match the size
of the HDFS blocks.

\noindent{\bf HDFS configuration.} We configured
io.file.buffer.size to be equal to the HDFS block size. The default value of
this parameter (64KB) results in too many sendFile system call, which in turn
create a lot of context-switching between user and kernel space, which results
in idleness for the IO device.  We modified the HDFS code to allow the
parameter to be set to 256MB as by default the maximum size is 32MB.

\subsection{Maximizing Device Throughput}
An important goal for HDFS is to maximize the throughput obtained from the
storage devices.  One way to achieve this is via multi-threading in the
DataNode. This is already part of the design as different DataNode threads can
serve different task requests concurrently.  While this can maximize device
throughput it does so at the expense of single-thread performance which reduces
task performance.

State-of-the-art data processing systems are heavily IO provisioned, with a
ratio of CPU to disk close 1~\cite{nutanix-platforms}. In this context, relying on parallelism to make
the most of the storage is unlikely to help because the number of tasks is
roughly the same as the number of disks (tasks are usually scheduled on a
separate core). As a result, it is important to understand and ensure that the
HDFS access pattern (single-thread, large files, sequential access, buffered
IO) can by itself extract maximum performance from SSDs.

\section{Methodology}
\label{sec:methodology}
In this section, we describe the hardware and software settings, tools,
workloads and metrics used in our analysis.

\subsection{Experimental Setup}
\textbf{Hardware.} We use three types of machines:
\textit{Machine A} has 2 Intel Xeon 2.4GHz E5-2630v3 processors, with 32 cores 
in total,  128GB of RAM, and a 450GB Intel DC S3500 Series (MLC) SATA 3.0 SSD.
\textit{Machine B} has 4 Intel Xeon 2.7GHz E5-4650 processors, with 32 cores 
in total, 1.5TB of RAM, and a 800GB HP 6G Enterprise SATA 3.0 SSD. 
\textit{Machine C} has 2 Intel Xeon 2.4GHz E5-2630v3 processors, with 32 cores 
in total, 128GB of RAM, and a 512GB Samsung 860 Pro (V-NAND) SATA 3.0 SSD.

Our SSDs have been very lightly used throughout their lifetimes. After
concluding our analysis we computed the total lifetime reads and writes
performed on the drives using the "sectors read" and "sectors written" fields
in /proc/diskstats in Linux. The value was less than 1TB for both reads and
writes for each drive. This is orders of magnitude less than the manufacturer
provided guarantees for SSDs. Thus, past heavy use of the drives is not a
factor in our findings. Moreover, the disk utilization of our SSDs in the
experiments is very low, under 20\%.

\textbf{Software.} We use Ubuntu 16.04 with Linux kernel version 4.4.0.
As a local file system we use ext4, one the most popular Linux file systems.
We use HDFS version 2.7.1.

\textbf{Monitoring Tools.} To monitor IO at the storage device, we rely on
block layer measurements using blktrace and blkparse. Blktrace collects IO
information at the block layer, while blkparse makes the traces human readable.
Where necessary we use perf and strace to analyze program behavior.

\textbf{Workloads.}
We use HDFS via Hadoop where we run a simple WordCount job. The exact type of
Hadoop job is inconsequential for our findings because we have already
decoupled HDFS performance from software bottlenecks in
Section~\ref{sec:decoupling}.  We modified the Hadoop WordCount job to not
write any output so that we can reliably measure read performance. We use the
FIO tool for analysis beyond HDFS. The data read by HDFS (or FIO) is composed
of randomly generated strings and is divided in 8 ext4 files of 256MB each. 

Our experiments consist in reading (with Hadoop or FIO) repeatedly over 24
hours the 8 ext4 files We ensure that all experiments run in isolation and are
not affected by interference from any other processes. 

\textbf{Presenting slowdowns.} For every slowdown we are able to separate its
effect and present results for periods with and without that slowdown. The
results without a slowdown include the effects of all other slowdowns that
occur at shorter timescales. For example, when comparing results with or
without temporal slowdown, the results include the effect of intrinsic slowdown
but not that of permanent slowdown. This is ok because the effect of a slowdown
is roughly constant over longer periods of time.

\textbf{What we measure.}
We measure performance at the HDFS DataNode level. We measure the throughput of
HDFS reads and the latency of individual block layer IO requests. We do not
measure end-to-end performance related to Hadoop tasks. Before every experiment
we drop all caches to ensure reads actually come from the drive. 

The Hadoop tasks are collocated with the HDFS DataNodes on the same machines.
The DataNodes send data to the tasks via the loopback interface using the
sendfile system call. We also analyzed short-circuit reads which enable Hadoop
tasks to read local input directly (using standard read calls) by completely
bypassing HDFS but the findings remained the same.

\textbf{Generating single requests.}
In our experiments using buffered IO, two requests overlap in the device
driver. In such a case, increases in request latency could be caused either by
drive internals or by a sub-optimal request overlap. To distinguish such cases
we tweak buffered IO to send one request at a time. To send a single request of
size X we first set the Linux read ahead size to X KB by tuning
\texttt{/sys/block/\textless device\textgreater/queue/read\_ahead\_kb}. 
We then use the \texttt{dd} command to read
one chunk of size X (dd bs=X count=1). The influence of read ahead size on
block layer size is known and discussed in related work~\cite{he2017unwritten}.

\subsection{Metrics}

In the rest of the paper, the word "file" refers to one 256MB ext4 file. In HDFS 
parlance this represents one HDFS block.

\textbf{File throughput.} The number of bytes, read from the target file, divided by
the period of time. The period starts with the submission of the first block
layer IO request in the file (as timestamped by blktrace) and finishes with the
completion of the last block layer IO request in the file (as timestamped by
blktrace). During this period we only count time when the disk is active, i.e.
there is at least one IO request being serviced by or queued for the drive. 
This metric removes the impact of disk idle time caused by context-switches 
between user and kernel space in the application. Our HDFS results show no
disk idle time after applying the changes in Section~\ref{sec:decoupling}.
Nevertheless, disk idle time appears in FIO and we chose to discard it for a fair
comparison to HDFS. Overall, the disk idle time does not influence our main findings.

\textbf{Request Latency.} The time between the timestamp when a block layer
request is sent to the drive (D symbol in blktrace) and the timestamp of its
completion (C symbol in blktrace). Both timestamps are taken from blktrace.

\textbf{Fragmentation.} The number of extents an ext4 file has. Note that all
of our files are 256MB. The maximum extent size in ext4 is 128MB. Therefore, the
minimum possible number of extents in a file is 2. 

\textbf{Recovery Time.} The period of time during which IO requests have higher
than usual latency due to intrinsic slowdown. This is measured starting from
the first IO read request of an ext4 extent until either the latency of the
requests decreases to normal or the extent is fully read, whichever comes first.

\section{Intrinsic Slowdown} \label{sec:IntSlwdown}

In this section, we introduce the intrinsic slowdown, a performance degradation
that predictably affects files at short time scales (seconds to minutes). This
slowdown is related to the logical file fragmentation. Every time a new file
system extent is read, a number of IO requests from the start of the extent are
served with increased latency and that is correlated with throughput drops.
Interestingly, even un-fragmented files are affected since a file has to have
at least one extent and every extent is affected by the slowdown. The more
fragmented a file is, the more extents it has, and the bigger is the impact of
the slowdown. 

Intrinsic slowdown appears on all the machines we tested and causes a drop in
throughput of 10-15\% depending on the machine. The slowdown lasts a variable
amount of time but there is no correlation with extent size. This slowdown
affects not only HDFS but all applications using buffered IO. 

The remainder of this section presents an overview of a throughput loss, an
analysis of the results, a discussion on causes and an analysis of mitigation
strategies.

\subsection{Performance Degradation at a Glance}\label{ssec:intMeasure}
\begin{figure}[t]
	\centering
	\includegraphics[width=\linewidth]{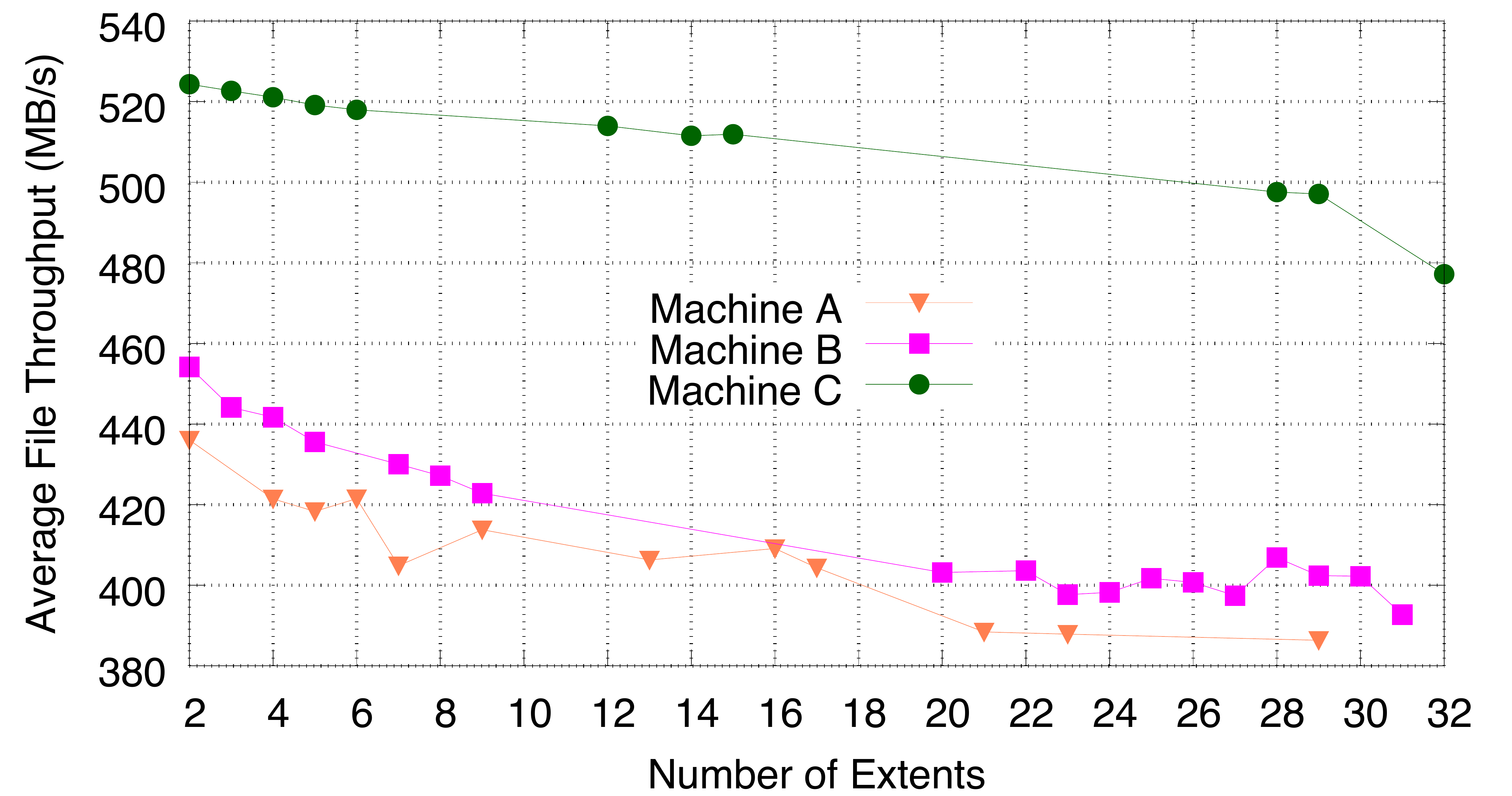}
	\caption{HDFS read. Average file throughput vs number of extents.}
	\label{fig:avg_thpt_vs_extents} 
	\vspace{-.2in}
\end{figure}

Figure~\ref{fig:avg_thpt_vs_extents} illustrates the influence that an
increased number of extents has on throughput for each of the 3 machines. In
this figure, each point represents the average file throughput of a set of
files with the same number of extents in one machine. Files were created using
ext4's default file allocation policies so we had no control over the number of
extent each file was allocated. We observed that ext4 allocations result in
highly variable fragmentation levels even on our drives which were less then
20\% full. We often saw cases where one file was allocated 30 extents and a
file created seconds after was allocated 2 extents. A thorough analysis of ext4
allocation patterns is, however, beyond the scope of this work.

The figure shows that an increase in fragmentation is correlated with a loss in
throughput. This finding holds on all 3 machines but the magnitude of the
throughput loss is different because the SSDs are different. With 29 extents,
throughput drops by roughly 13\% for machines A and B, but by less than 5\% for
machine C. There is a limit to the throughput loss and that is best exemplified
by the fact that throughput drops very slowly for machines A and B after 20
extents. The reason is that the extents are smaller but the recovery period is
not correlated with the extent size so a very large percentage of the IO
requests is affected by the slowdown. 

\subsection{Analysis}\label{ssec:intAnalysis}

\begin{figure}[t]
	\centering
	\includegraphics[width=\linewidth]{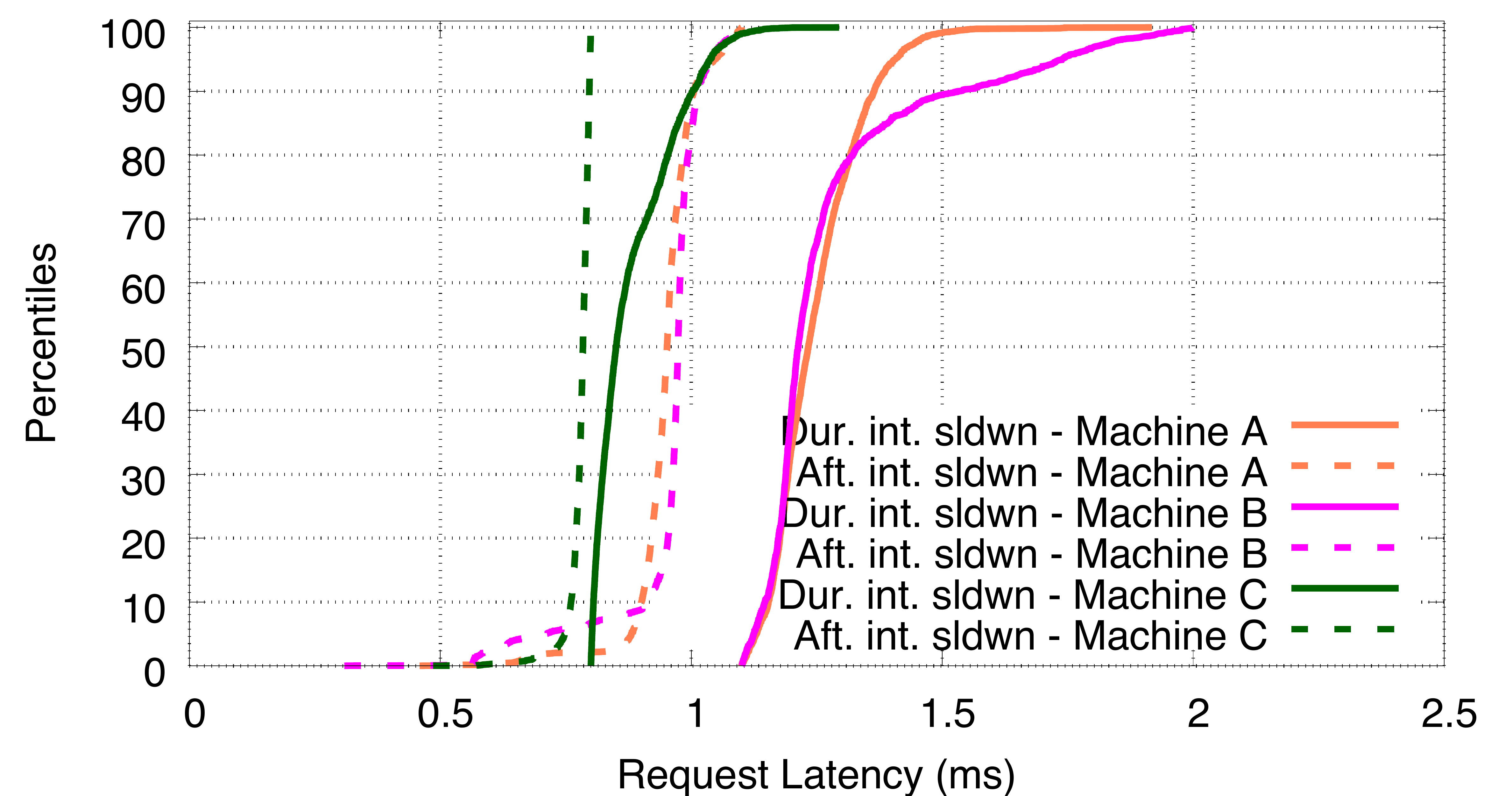}
	\caption{HDFS read. Request Latency CDFs during and after intrinsic slowdown.}
	\label{fig:cdf_lat_all_intrinsic} 
	\vspace{-.2in}
\end{figure}

\textbf{Correlations.} We next analyze IO request latency.
Figure~\ref{fig:cdf_lat_all_intrinsic} presents the request latencies on
machines A, B and C, during an HDFS read. The dashed lines correspond to
request latencies after the intrinsic slowdown disappeared while the continuous
one show latencies during the slowdown. Latencies increase during slowdown both
at the median but especially in the tail. Machine C shows both the smallest
latencies and the smallest degradation and this is due to the fact that its SSD
is based on a different technology (V-NAND). 

The above latencies are with the standard buffered IO configuration in which 2
requests overlap in the device driver. We also measured (not illustrated)
the request latency when sending one single request a time. We find that
latency remains unaffected even during the parts of the extent that are
normally affected by intrinsic slowdown. This suggests that the latency
increase and throughput loss are not solely due to drive internals.  This is
expected as the SSD FTL has no notion of extents which are a file system
construct. 

\begin{figure}[t] \centering
\includegraphics[width=\linewidth]{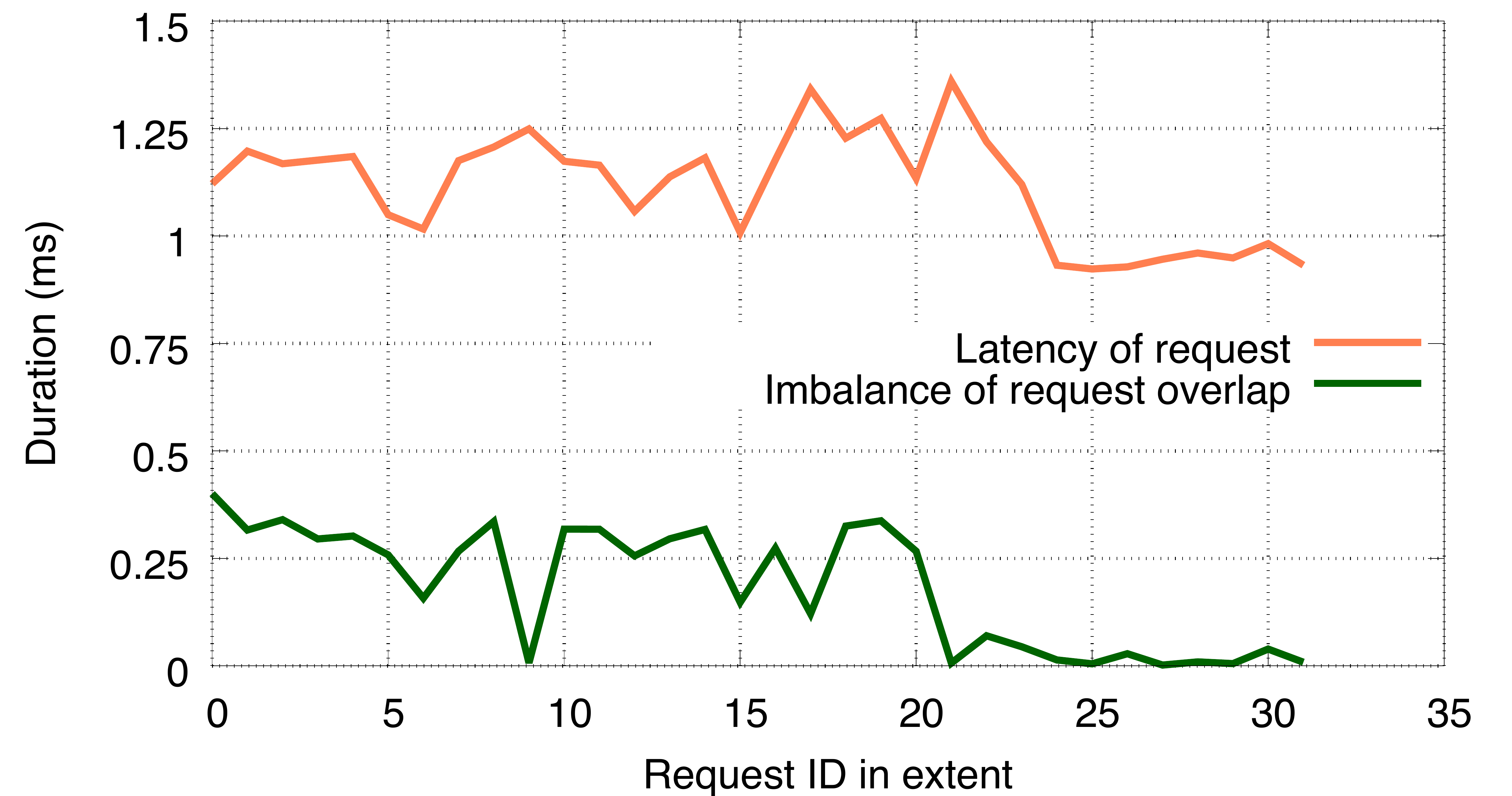}
\caption{Correlation between increased latency and request overlap imbalance.}
\label{fig:correlation_overlap_imbalance} \vspace{-.2in} \end{figure}

We find that inherent slowdown is correlated with a sub-optimal request
overlap in the device.  Consider a request $R_2$ and let $S_2$ and $E_2$ be its
en-queueing and completion time. With buffered IO, the execution of $R_2$ overlaps
with the final part of $R_1$ and the first part of $R_3$, $R_1$ being the
previous request and $R_3$ the next. We have that $S_2 < E_1 < S_3 < E_2$.  We
find that periods of intrinsic slowdown are correlated with an imbalanced
overlap, that is $R_2$ overlaps much more with either $R_1$ or $R_3$. In other
words, imbalance overlap occurs when $T >> 0$  where
$T=abs((E_2-S_3)-(E_1-S_2))$.  To exemplify,
Figure~\ref{fig:correlation_overlap_imbalance} shows the correlation between
$T$ and request latency for a sample extent. For the first 20 requests, the
overlap is sub-optimal and latency suffers. The overlap imbalance is corrected
around request 22 and soon after latency drops under 1ms which is the latency
we normally see outside of intrinsic slowdown.
  
\textbf{Characterization of recovery periods} 
We next analyze the duration and variability of the recovery periods.  There are
two main insights. First, even for a single extent size and one machine, there
can be significant variation in the duration of the recovery period.  Second,
the duration of the recovery period is not correlated to the extent size.

\begin{figure}[t]
	\centering
	\includegraphics[width=\linewidth]{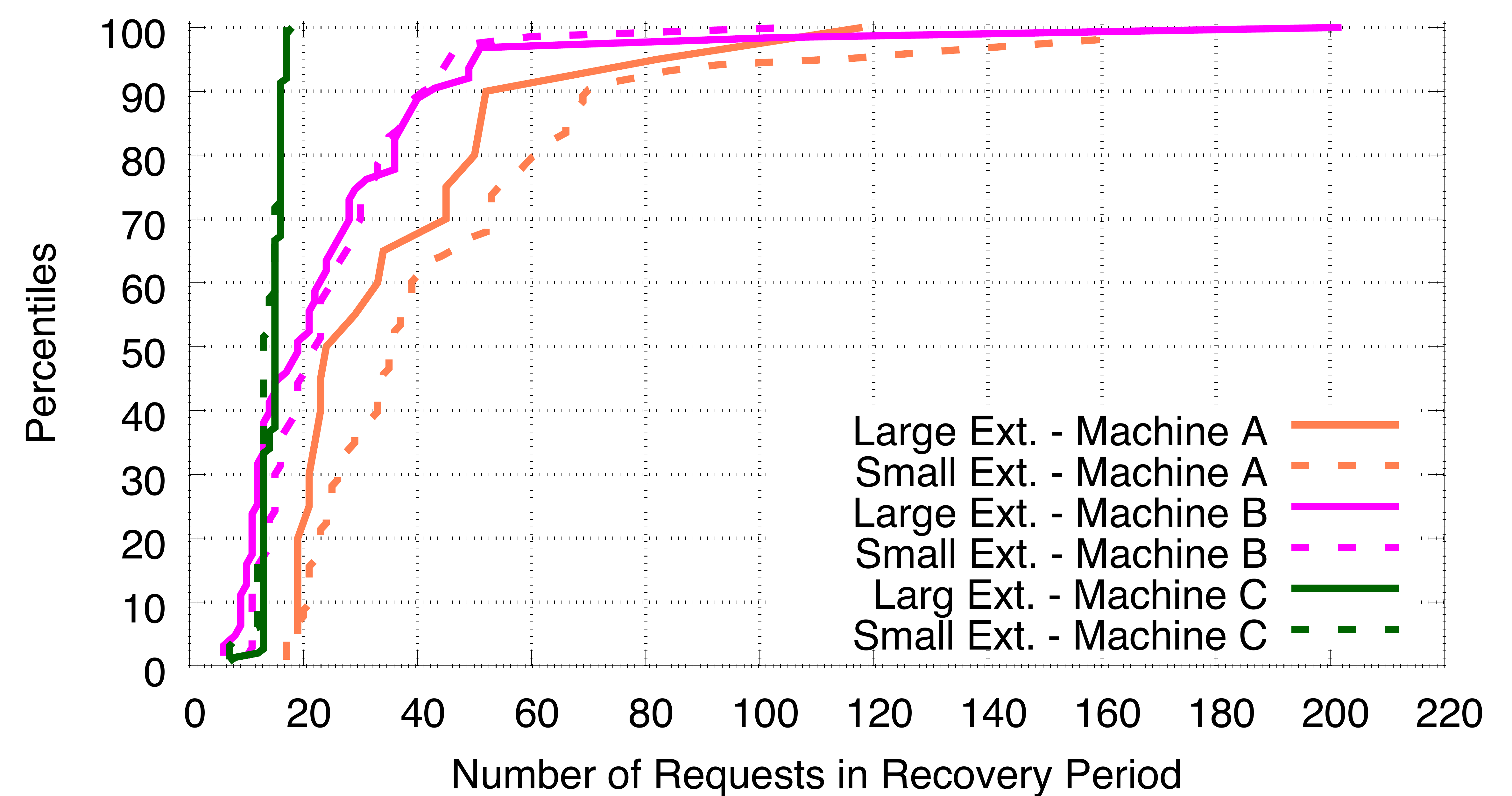}
	\caption{HDFS read. CDFs of number of requests executed in the recovery period for
	large and small extents.}
	\label{fig:cdf_recovery_intrinsic} 
	\vspace{-.2in}
\end{figure}

Figure~\ref{fig:cdf_recovery_intrinsic} shows CDFs of the duration of the
recovery period on the 3 machines. For each machine we show large extents
(128MB) with continuous lines and smaller extents (32-40 MB) with dashed lines.
We aggregated results from extents from multiple files if they have the
target size and reside on the same machine. We measure the recovery duration in
number of requests. The request size is 256KB. 

The CDFs for any one of the machines show a similar pattern in the recovery period
despite the different extent size. Therefore, extent size is not a factor with
respect to the duration of the recovery period.

There is significant variability in the recovery period for every extent size
on machines A and B. The worst-case recovery duration is more than 5x that of
the best-case. In contrast, machine C shows much less variability.

If we compute the recovery period relative to extent size (not illustrated) we
find that for the smallest extents (e.g. 8MB) it is common for at least 50\% of
the requests in the extent to be affected by intrinsic slowdown. In the worst
case, we have seen 90\% of an extent being affected.

\noindent{\bf Discussion on internal SSD root cause.\hspace{0.2in}}
Since we do not have access to the proprietary SSD FTL design we cannot
directly search for the root cause internal to the drive. We believe that
sub-optimal request overlap leads to throughput loss because it forces the
drive to be inefficient by serving both overlapping requests in parallel when
the most efficient strategy would sometimes be to focus on the oldest one
first. The request stream enters in this state due to the initial requests at
the start of the extent. The stream self-corrects by eventually reaching the
optimal (balanced) request overlap and remaining there. The software does not
help in the correction as it functions in a reactive manner. It sends a new
request as soon as one completes. The self-correction happens solely due to
timing, based on the request latencies. This also explain the variability in
the recovery periods.

\subsection{Mitigation Strategies}\label{ssec:intMitigatoin}

We consider mitigation strategies that are more aggressive in generating
request level parallelism in the hope that they could compensate for the loss
in throughput due to the slowdown. We find that both direct IO as well as
increasing the number of requests sent in parallel with buffered IO can mask
intrinsic slowdown.

\begin{figure}[t]
	\centering
	\includegraphics[width=\linewidth]{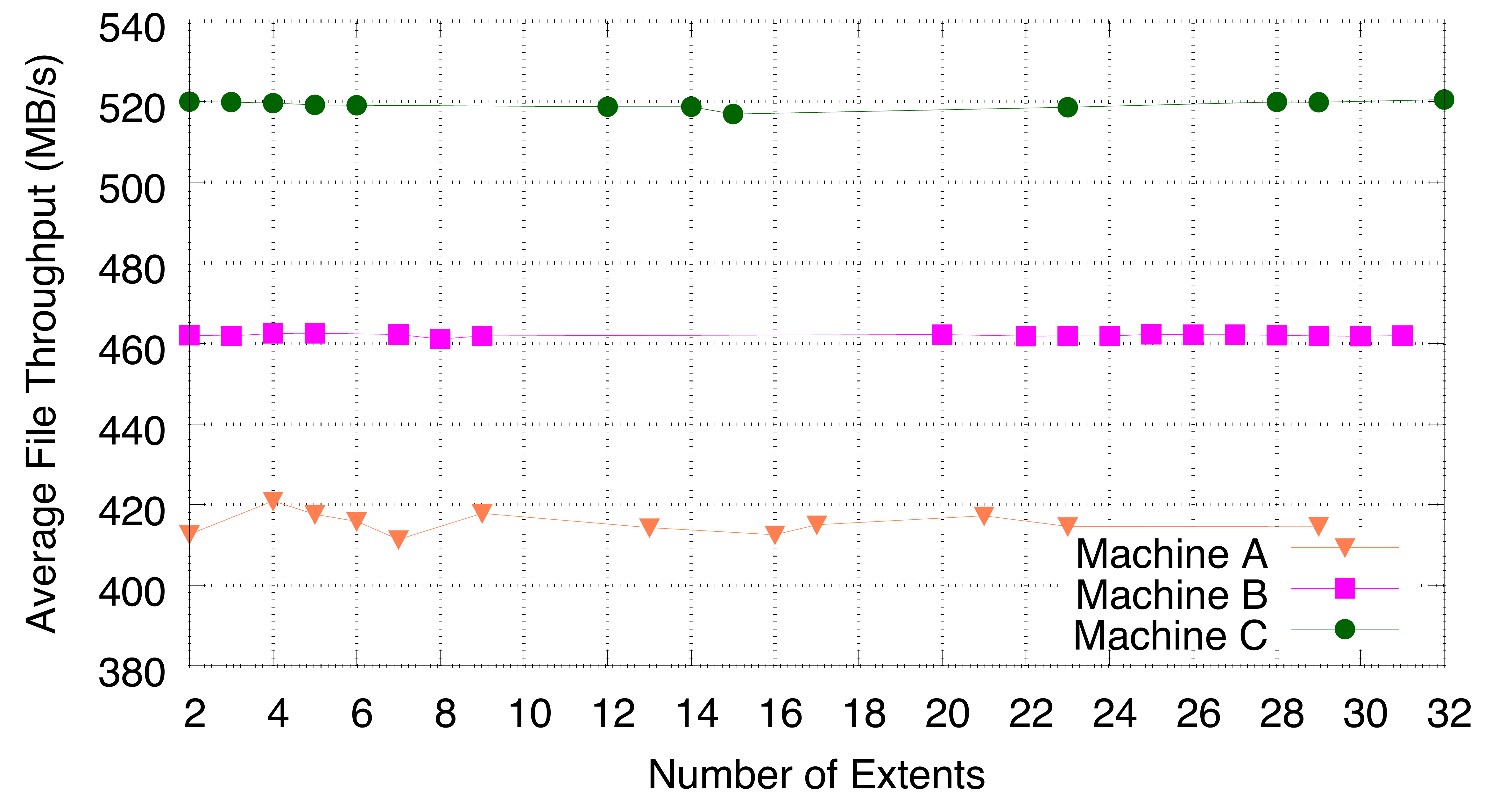}
	\caption{FIO Direct IO read. Average file throughput vs number of extents.}
	\label{fig:directIO_avg_thpt} 
	\vspace{-.2in}
\end{figure}

Figure~\ref{fig:directIO_avg_thpt} compares average file throughput vs number
of extents, when using direct IO across different machines. The files are the
same as in Figure~\ref{fig:avg_thpt_vs_extents}.  The figure shows that average
throughput is maintained across different numbers of extents with direct IO.
The tendency holds across all machines tested. In other words, direct IO can
mask intrinsic slowdown. The reason is that by sending more and larger
requests, direct IO better leverages the device parallelism. We observe the
same effect when increasing parallelism in buffered IO, by increasing the read
ahead size. This setting results in both larger requests as well as more
requests being sent in parallel to the drive.

\section{Temporal Slowdown}\label{sec:TmpSlwdown}

In this section, we introduce the temporal slowdown, a periodic and temporary
performance degradation that affects files at medium timescales (minutes to
hours). At the high level, the pattern in which temporal slowdown manifests
might we confused with write-induced SSD garbage collection (GC). However,
temporal slowdown is not always GC. Surprisingly, on machine A, it always
manifests even in read-only workloads. On machine B, it is indeed triggered by
writes but interestingly it takes a very small amount of writes relative to the
drive capacity to trigger temporal slowdown. Moreover, our SSDs have a very low
utilization (under 20\%). We link the slowdown to tail latency increases inside
the drive. Temporal slowdown causes a throughput drop of up to 14\%. Temporal
slowdown affects not only HDFS but all applications using either direct or
buffered IO. 

The remainder of this section presents an overview of a throughput loss, an
analysis of the results, a discussion on causes and an analysis of mitigation
strategies.

\subsection{Performance Degradation at a Glance}\label{ssec:tmpMeasure}

\begin{figure}[t]
	\centering
	\includegraphics[width=\linewidth]{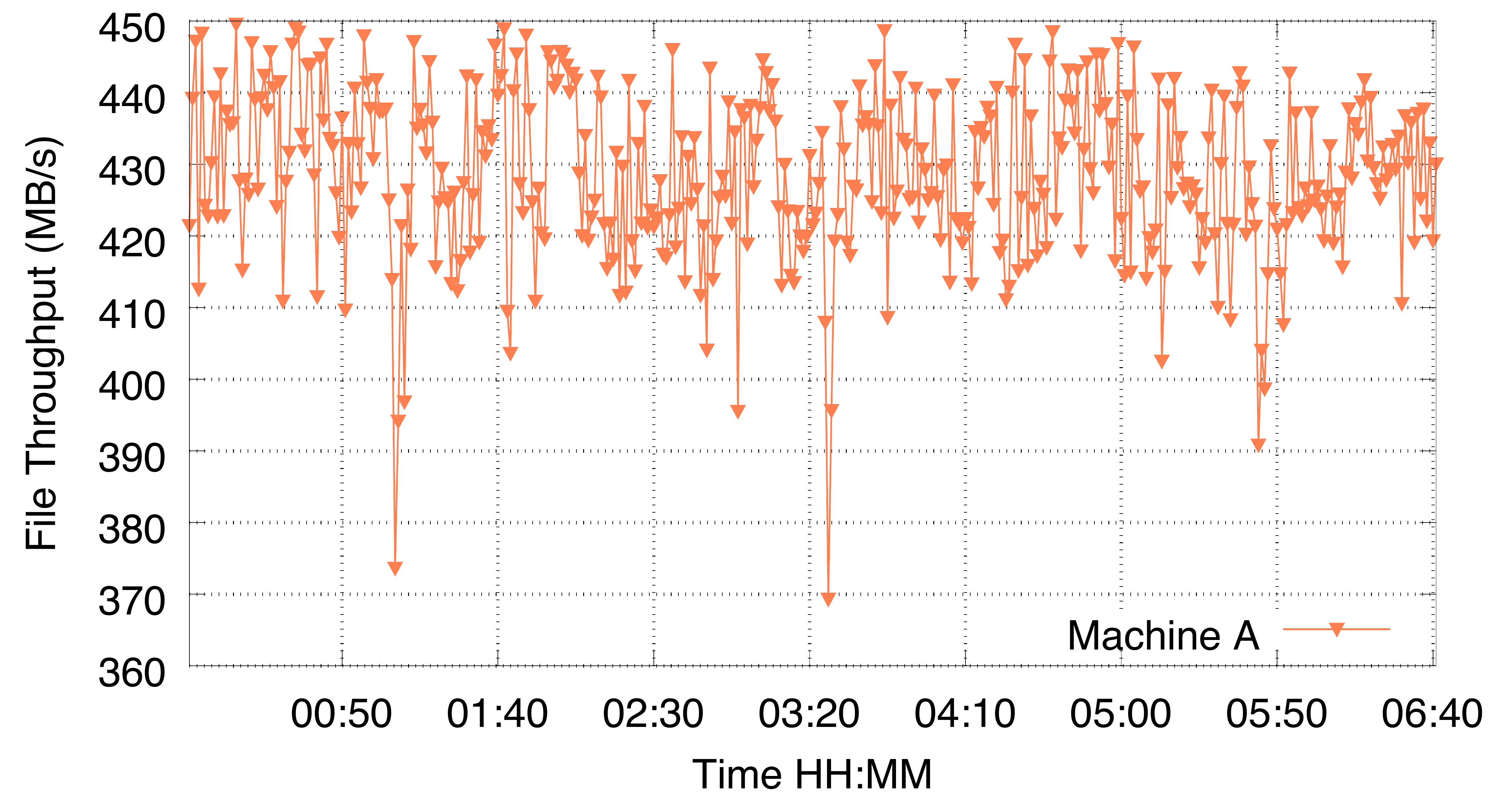}
	\caption{HDFS read. File throughput timeline on Machine A.}
	\label{fig:HDFS_trex_thpt_var} 
	\vspace{-.2in}
\end{figure}

Figure~\ref{fig:HDFS_trex_thpt_var} presents the throughput timeline of a file
affected by temporal slowdown on machine A. It shows three instances of
the slowdown around the 1:00, 3:30 and 5:40 marks. The rest of the
throughput variation is caused by inherent slowdown. The average throughput of
the periods not affected by the slowdown is 430 MB/s. The first instance of
slowdown causes a drop in throughput to 370 MB/s, a 14\% drop from the 430 MB/s
average. On machine A, temporal slowdown appears on average every 130 min
and last on average 5 min.

\begin{figure}[t]
	\centering
	\includegraphics[width=\linewidth]{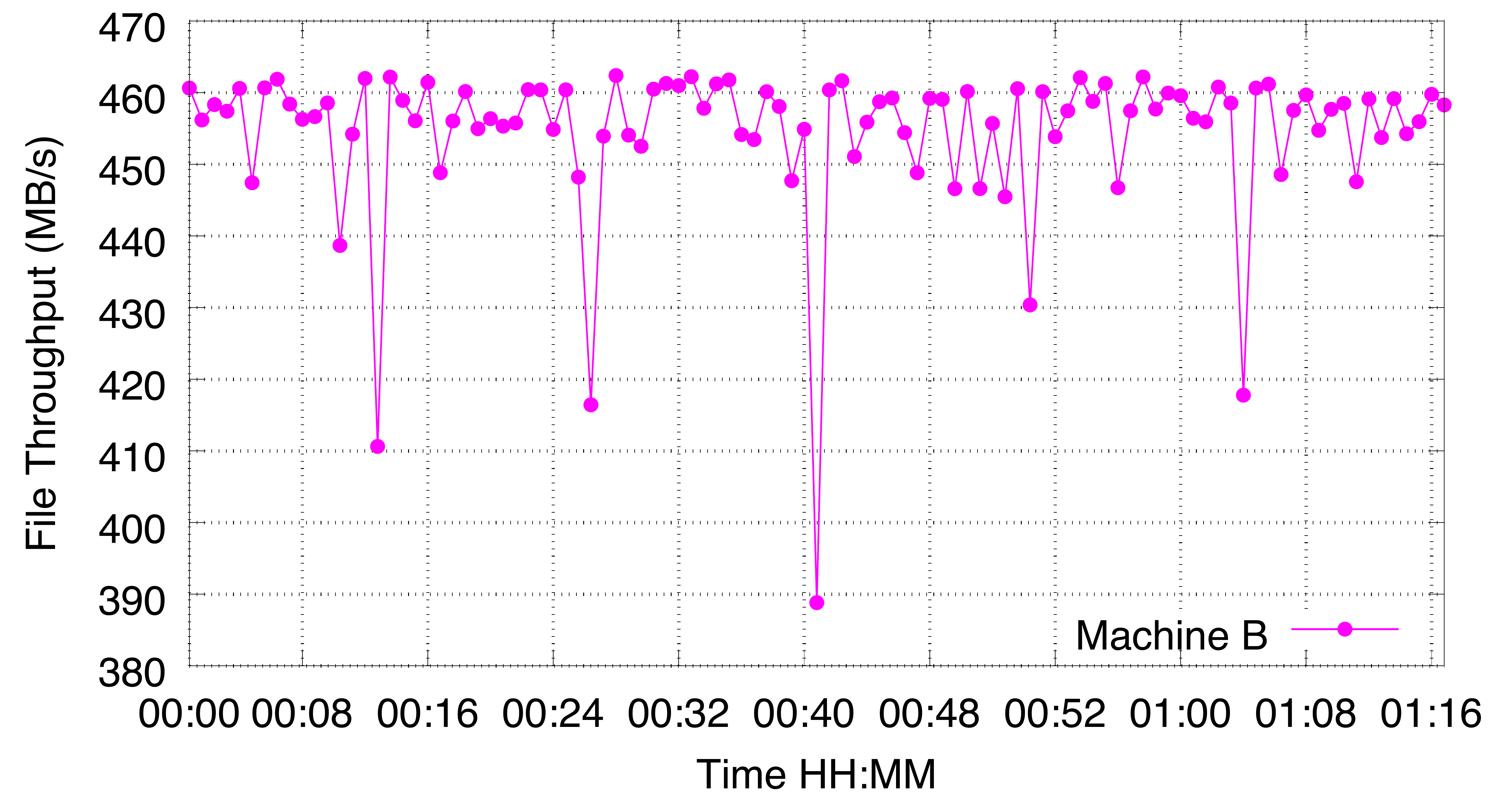}
	\caption{HDFS read. File throughput timeline on Machine B.}
	\label{fig:HDFS_fatnode_thpt_var} 
	\vspace{-.2in}
\end{figure}

Figure~\ref{fig:HDFS_fatnode_thpt_var} shows the same experiment on machine B.
There are 5 instances of temporal slowdown clearly visible due to the
pronounced drops in throughput. The average throughput of the periods not
affected by the slowdown is 455 MB/s. The biggest impact is caused by the third
slowdown instance which causes a drop to 390 MB/s, almost 15\% down from the
average. On machine B, temporal slowdown appears on average every 18 min
and last for 1.5 min.

\subsection{Analysis}\label{ssec:tmpAnalysis}

\begin{figure}[t]
	\centering
	\includegraphics[width=\linewidth]{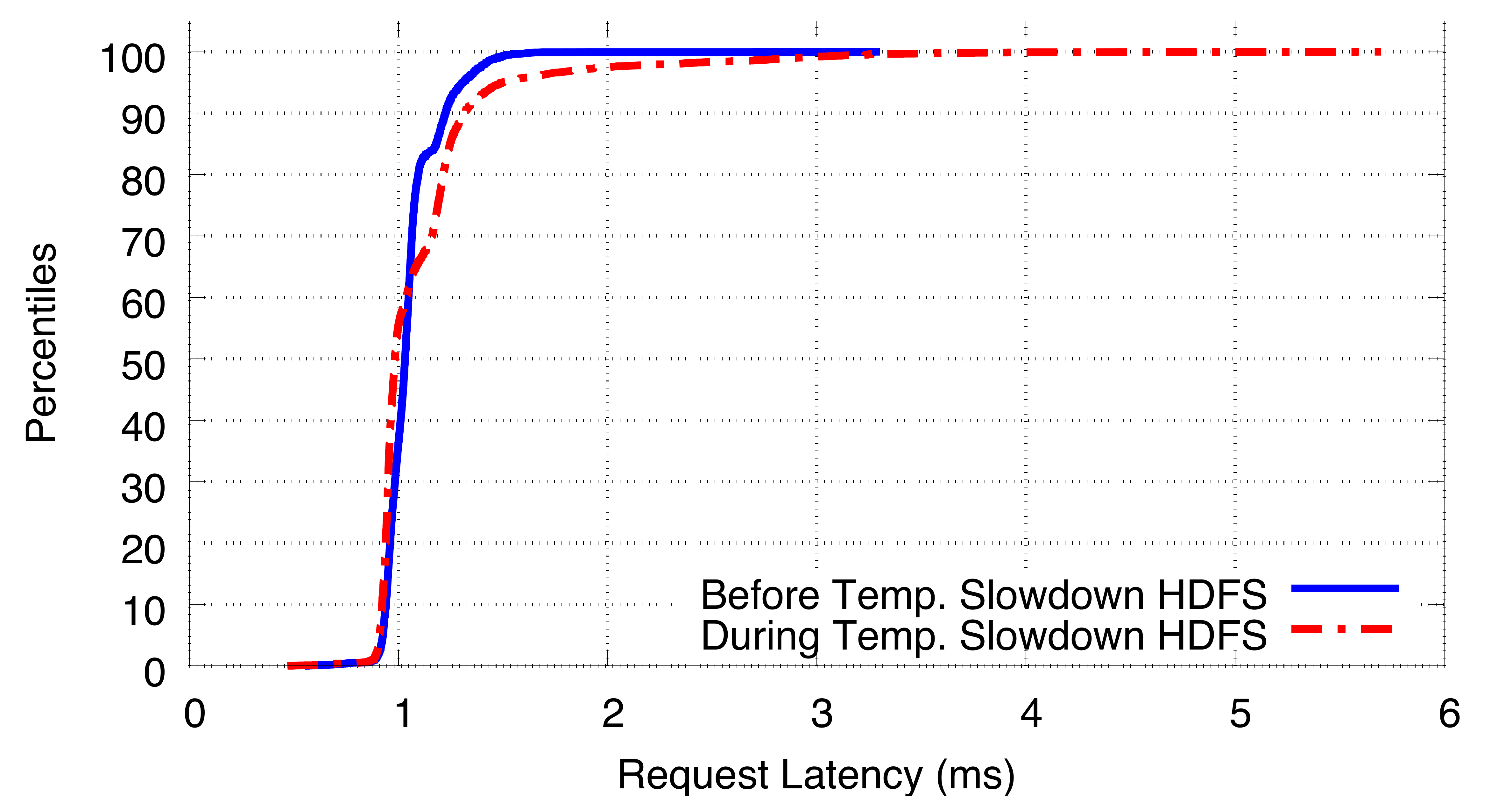}
	\caption{HDFS read. CDFs of read request latencies during and outside of temporal slowdown on Machine A.}
	\label{fig:HDFS_trex_latencies_cdf} 
	\vspace{-.2in}
\end{figure}

\noindent{\bf Correlations.\hspace{0.2in}} We next analyze IO
request latency.  Figure~\ref{fig:HDFS_trex_latencies_cdf} shows a CDF of the
request latencies for one file.  One line shows latencies during temporal
slowdown while another shows latencies during periods not affected by the
slowdown. The difference lies in the tail behavior. During temporal slowdown a
small percentage of the requests show much larger latency. This is consistent
with the impact of background activities internal to the drive.

The experiments in Figures~\ref{fig:HDFS_trex_thpt_var} and
\ref{fig:HDFS_fatnode_thpt_var} do introduce writes and they responsible for
triggering temporal slowdown on machine B. Even though from the application
perspective (i.e. Hadoop) the workload is read-only, a small number of writes
appear due to HDFS metadata management.  These are the only writes in the
system as we explicitly turned off journaling and metadata updates in ext4.
Interestingly, a small amount of writes relative to the drive size is
sufficient to trigger temporal slowdown.  On machine B, temporal slowdown
occurs approximately every 120MB. That amounts to only 0.015\% of the disk
size.

\noindent{\bf Temporal slowdown without writes.\hspace{0.2in}} Our main finding
related to temporal slowdown is that it can occur in the complete absence of
writes. This occurs only on machine A so we focus on it for these experiments.
To avoid any writes, we repeat the experiment using FIO instead of HDFS.  We
configure FIO to use the \texttt{read} system call and evaluated both direct IO
and buffered IO. The results were similar so we only show direct IO. We confirm
that there are no writes performed during the experiments by checking the
number of written sectors on the drive (from \texttt{/proc/diskstats}), before
and after the experiments. In addition, we ensure that no writes have been
performed in the system for at least one hour before the start of the
experiments.

\begin{figure}[t]
	\centering
	\includegraphics[width=\linewidth]{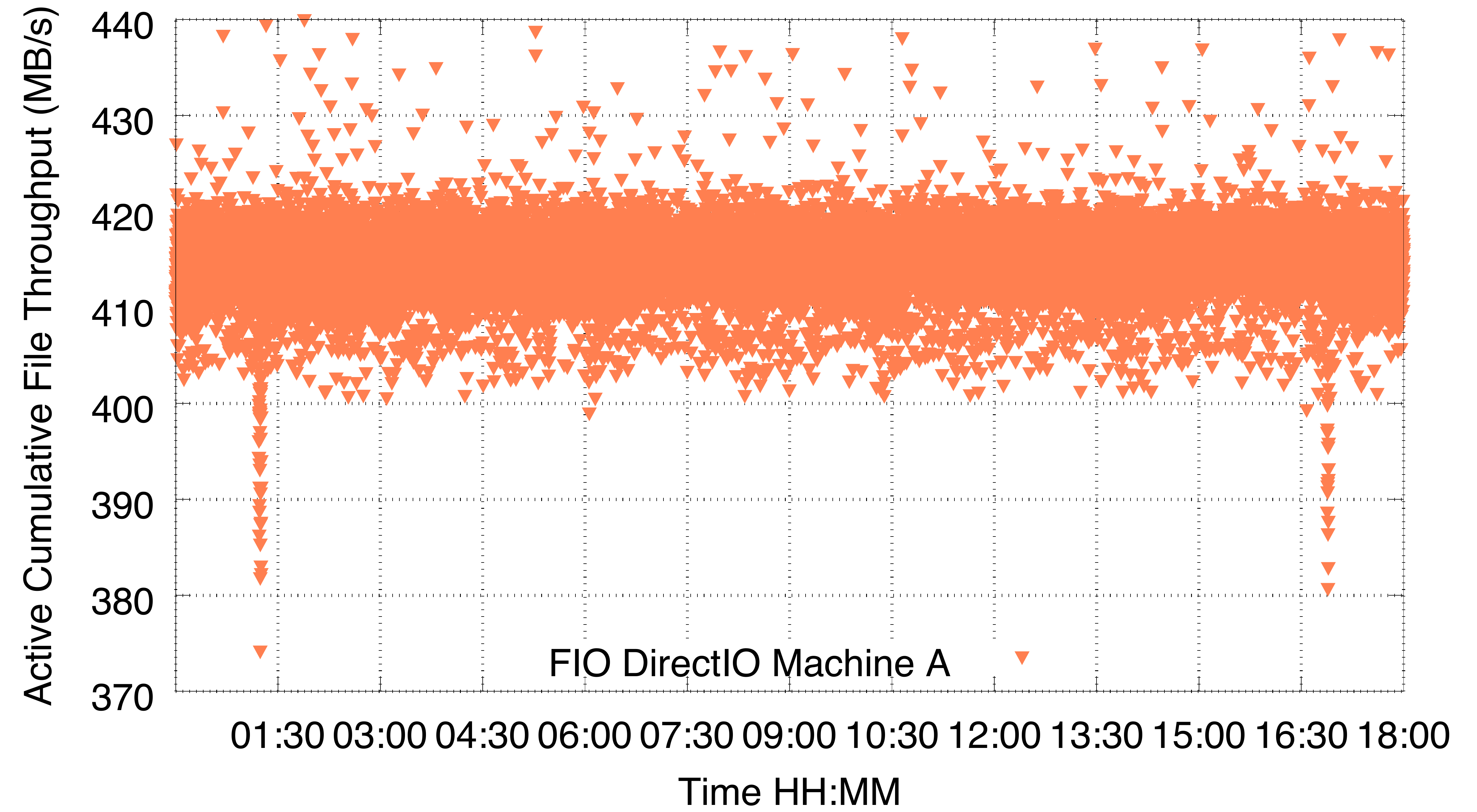}
	\caption{FIO direct IO Read. File throughput timeline on Machine A.}
	\label{fig:FIO_directIO_trex_thpt} 
	\vspace{-.2in}
\end{figure}

In Figure~\ref{fig:FIO_directIO_trex_thpt}, we show the throughput timeline
when using FIO with direct IO. FIO shows more variability in the common case
compared to Hadoop because of context-switches between kernel and user space.
The temporal slowdown is again visible despite the absence of writes. The
slowdown appears every 130 min on average and last 5 min on average. The
periodicity is almost identical to the HDFS case, suggesting that the HDFS
metadata writes did not play a role in triggering temporal slowdown on machine
A.

\begin{figure}[t]
	\centering
	\includegraphics[width=\linewidth]{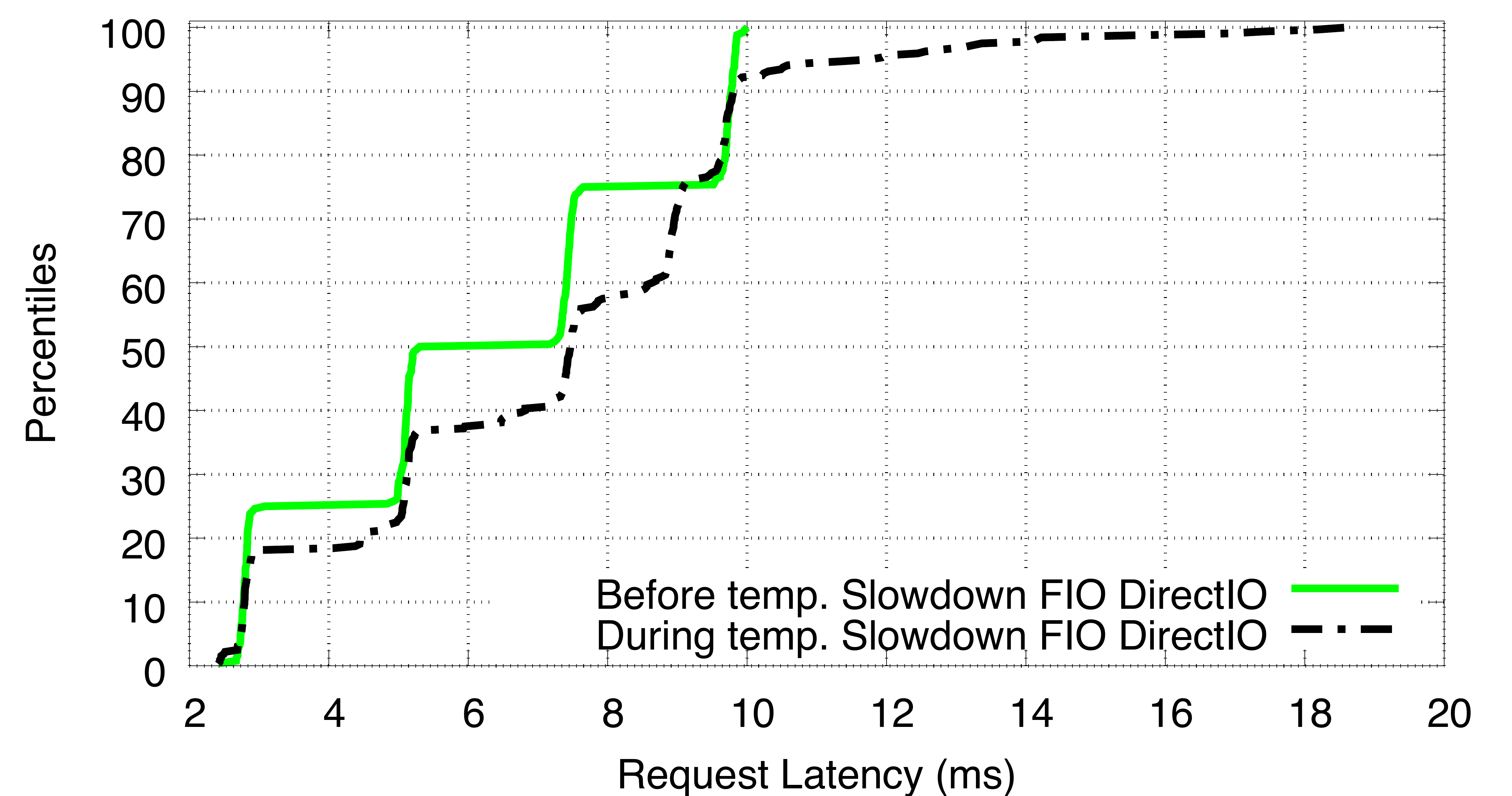}
	\caption{FIO Direct I/O Read. CDFs of read request latencies before
	and during temporal slowdown on MachineA.}
	\label{fig:FIO_direct_trex_latencies_cdf} 
	\vspace{-.2in}
\end{figure}

Figure~\ref{fig:FIO_direct_trex_latencies_cdf} presents the IO request latency
for FIO with direct IO. Again, tail latency increases during slowdown. The four different
latency steps appear because direct IO sends, by default, four large requests (1 MB) to
the drive.

\noindent{\bf Trigger of slowdown without writes.\hspace{0.2in}}
Next, we analyze whether temporal slowdown in the absence of writes is
correlated with the number of reads performed or it is time-based. We introduce
periods of inactivity using sleep periods between the reads.  We make sure
that these periods are much smaller than the duration of temporal slowdown so
that we do not miss slowdown events. We find that regardless of the inactivity
period induced, the periodicity remains the same suggesting time-based
triggers.

\noindent{\bf Discussion on internal SSD root cause.\hspace{0.2in}}
Since we do not have access to the proprietary SSD FTL design we cannot
directly search for the root cause internal to the drive.  In theory, there are
three known culprits for temporal slowdowns in SSDs yet our findings does not
match any of them. The first one is write-induced
GC~\cite{skourtis2014flash,yan2017tiny}.  However, we show that temporal
slowdown can appear in the absence of writes as well. The last two culprits are
read disturbance and retention errors~\cite{mutlu_errors_date12}. In the
related work, in Section \ref{sec:related}, we argue at length that these
culprits appear on drives that are far more worn out (orders of magnitude more
P/E cycles) than ours and after order of magnitude more reads have been
performed.  We hypothesize that temporal slowdown on our drives is triggered by
periodic internal bookkeeping tasks unrelated to past drive usage or current
workload.

\subsection{Mitigation Strategies}\label{ssec:tmpMitigation}
We have found no simple way of masking temporal slowdown. It occurrs for both
buffered IO and direct IO. One could attempt to detect early signs of slowdown
or estimate its start via profiling and then avoid performing reads during the
period. This would yield more predictable performance at the expense of
delays.

\section{Permanent Slowdown}\label{sec:PermSlwdown}

In this section, we introduce the permanent slowdown, an irreversible
performance degradation that affects files at long timescales (days to weeks).
Permanent slowdown occurs at a \textit{file level}. It is not triggered by a
single drive-wide event. Thus, at any point in time, a drive can contain both files
affected by permanent slowdown and files unaffected by it. The exact amount of
time it takes for a file to be affected by permanent slowdown varies from file
to file and is not influenced by how many times a file was read.  We only see
permanent slowdown on machines of type A.  Permanent slowdown causes a
throughput drop of up to 15\%. 

We find that permanent slowdown is not specific to HDFS but affects all read
system call that use buffered IO. We link the slowdown to unexpected and
permanent latency increases inside the drive for all IO requests. 

For terminology, in the context of permanent slowdown, "before" means before
the first signs of slowdown and "after" means after slowdown completely set in.
The CDFs represent a single HDFS file composed of 8 blocks (i.e. 8 ext4 files).
Figure~\ref{fig:HDFS_thpt_drop} shows a different file where we caught the
onset of the slowdown.  Nevertheless, we have seen that all files affected by
the slowdown show a similar degradation pattern and magnitude.

The remainder of this section presents an overview of a throughput loss, an
analysis of the results, a discussion on causes and an analysis of mitigation
strategies.

\subsection{Performance Degradation at a Glance}\label{ssec:permMeasure}
\begin{figure}[t]
	\centering
	\includegraphics[width=\linewidth]{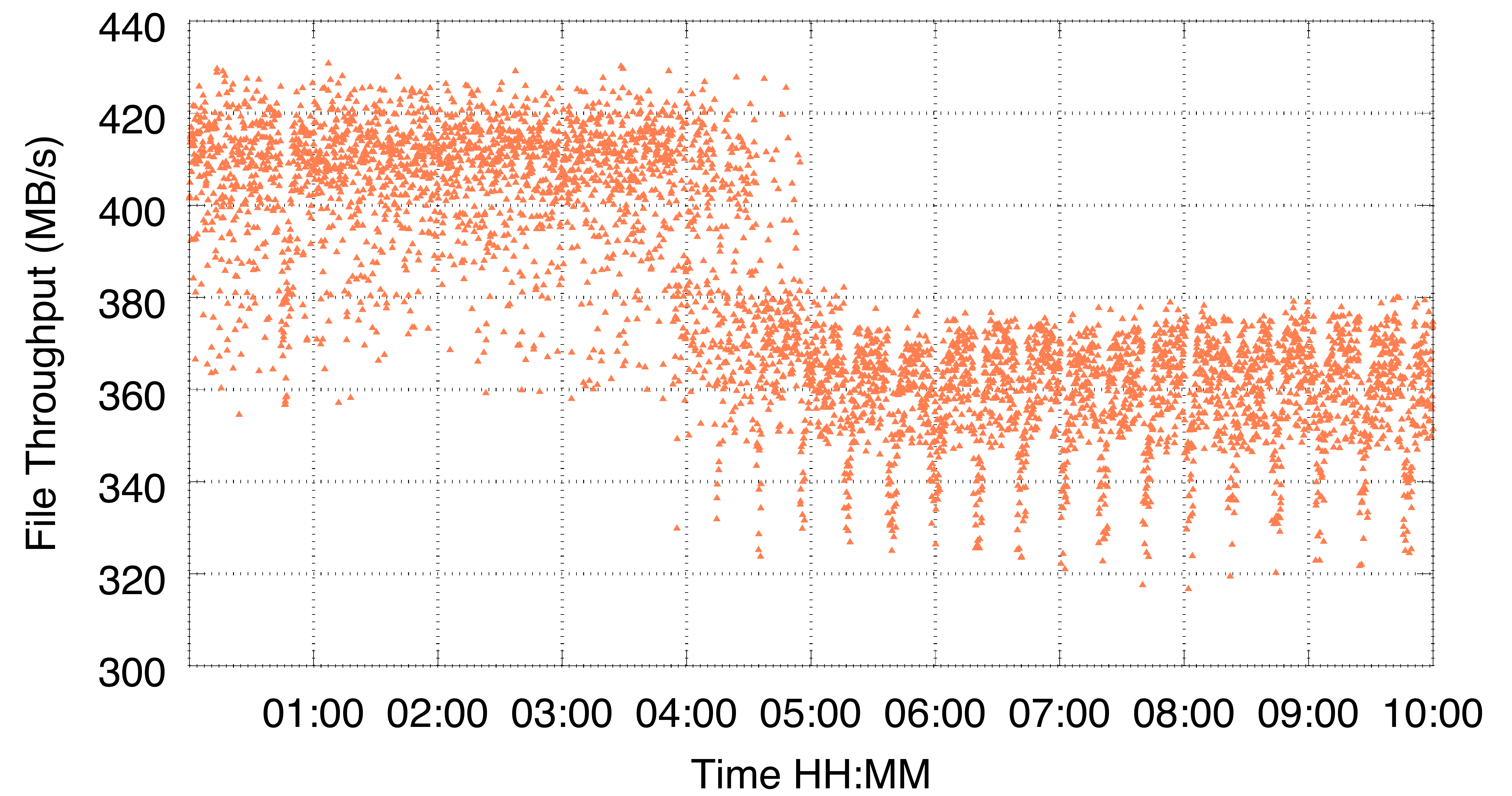}
	\caption{HDFS read. File throughput over a 10h period centered around the onset of permanent slowdown.}
	\label{fig:HDFS_thpt_drop} 
	\vspace{-.2in}
\end{figure}

Figure~\ref{fig:HDFS_thpt_drop} shows the onset and impact of permanent
slowdown. The plot shows a 10 hour interval centered around the onset of
permanent slowdown. The file was created several days before this experiment
was ran. For the first 4 hours, read throughput lies between 340 MB/s and 430
MB/s. This variation is explained by the intrinsic and the temporal slowdowns
described in Sections~\ref{sec:IntSlwdown} and \ref{sec:TmpSlwdown}. Around the
fourth hour, the permanent slowdown appears and after less than one hour it
completely sets in. From that point on, read throughput remains between 320
MB/s and 380 MB/s in this experiment and all future experiments involving this
file. 

\begin{figure}[t]
	\centering
	\includegraphics[width=\linewidth]{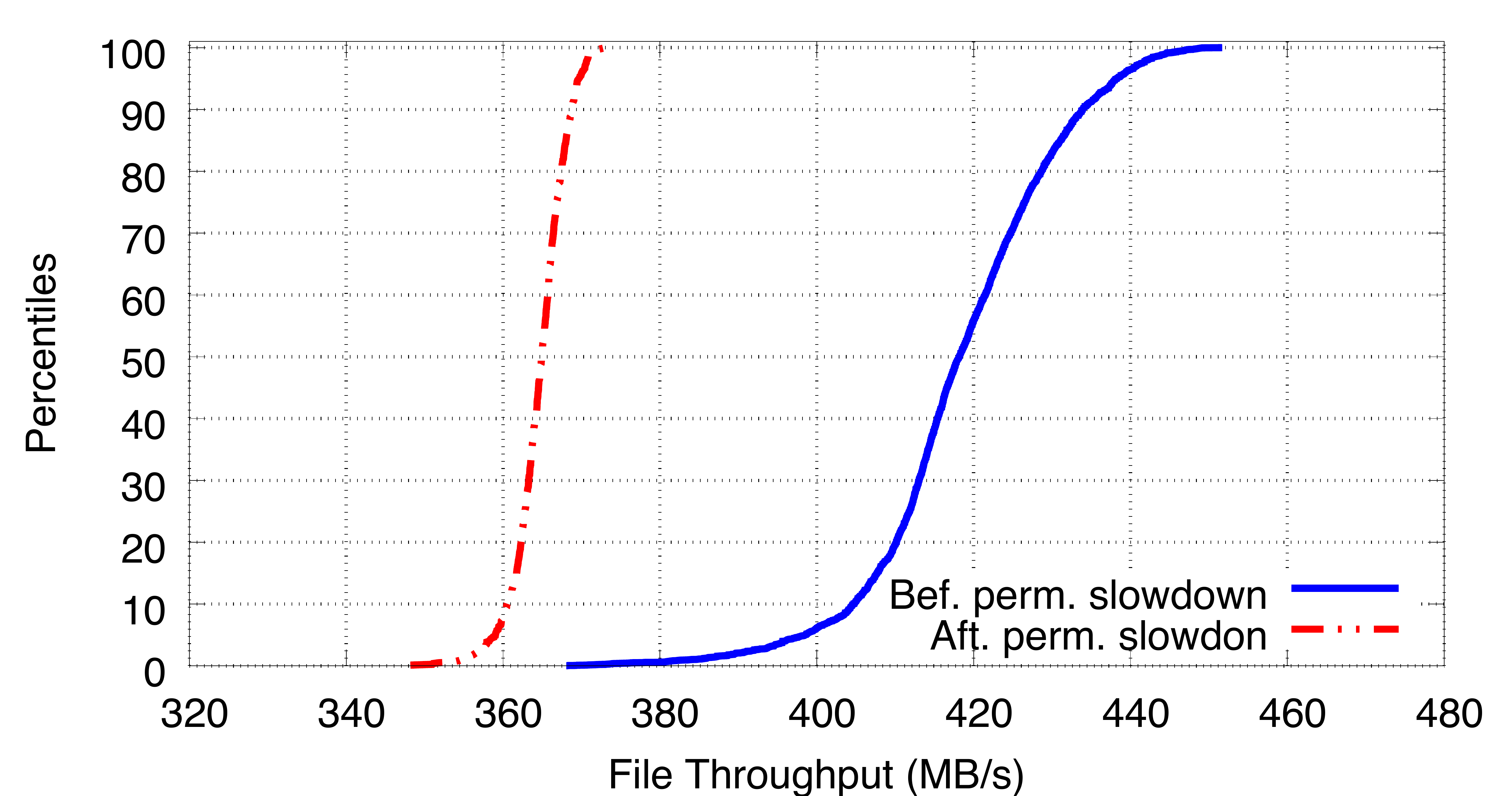}
	\caption{HDFS read. CDFs of file throughput before and after permanent slowdown.}
	\label{fig:HDFS_cdf_thpt} 
	\vspace{-.2in}
\end{figure}

Figure~\ref{fig:HDFS_cdf_thpt} compares the CDF of the read throughput of the
same file before and after slowdown. At the median, throughput drops
by 14.7\% from 418 MB/s to 365 MB/s.

\subsection{Analysis}\label{ssec:permAnalysis}

\noindent{\bf Generality.\hspace{0.2in}} We start by analyzing the generality
of the permanent slowdown. HDFS uses the sendfile system call to transfer data.
Using the \texttt{perf} tool we find that sendfile shares most of its IO path
in the Linux kernel with the read system calls that use buffered IO. Therefore,
we ask whether permanent slowdown affects only sendfile system calls or also
read system calls that use buffered IO.

\begin{figure}[t]
	\centering
	\includegraphics[width=\linewidth]{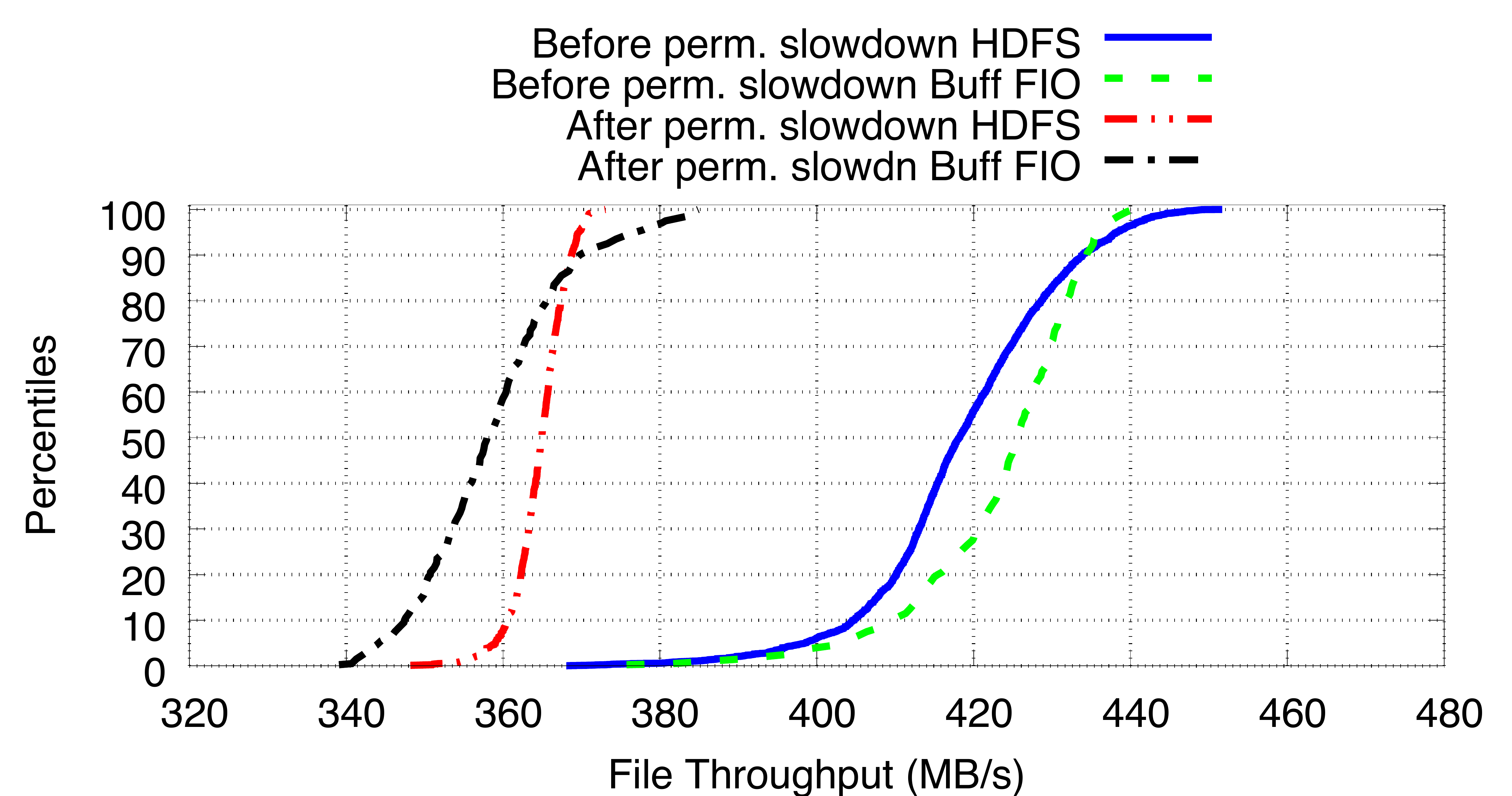}
	\caption{HDFS read vs Buffered FIO. CDFs of file throughput before and after permanent slowdown.}
	\label{fig:cdf_thpt_hdfs_fio_buff} 
	\vspace{-.2in}
\end{figure}

We use FIO to generate reads using buffered IO. We configure FIO to use the
read system call (i.e. sync io engine as a FIO parameter).
Figure~\ref{fig:cdf_thpt_hdfs_fio_buff} presents a throughput comparison
between HDFS (sendfile system call) and FIO (read system call). The two
rightmost CDFs show the throughput for HDFS and FIO before permanent slowdown.
HDFS and FIO behave similarly. The same applies after permanent slowdown sets
in (leftmost CDFs). Similar results were obtained using libaio as an IO engine
for FIO. This result show that permanent slowdown does not affect a particular
system call (sendfile) but the group of read system calls that perform buffered
IO.

\begin{figure}[t]
	\centering
	\includegraphics[width=\linewidth]{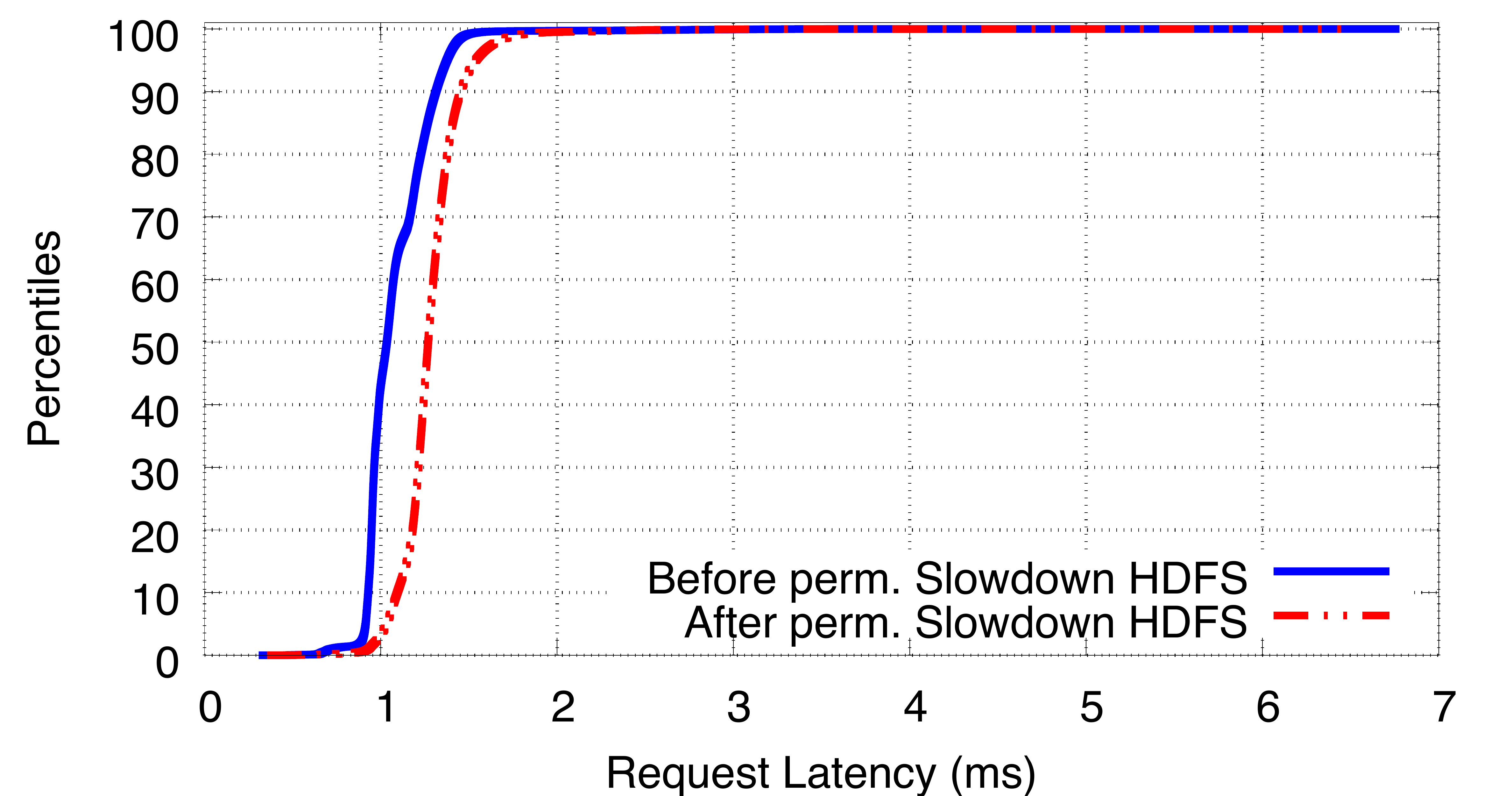}
	\vspace*{-7mm}
	\caption{HDFS read. CDFs of request latency before and after permanent slowdown.}
	\label{fig:HDFS_cdf_lat} 
	\vspace{-.2in}
\end{figure}

\noindent{\bf Correlations.\hspace{0.2in}} We next analyze IO request latency.
Figure~\ref{fig:HDFS_cdf_lat} compares the CDFs of request latency in HDFS on
one file before and after permanent slowdown. Permanent slowdown induced an
increase in latency at almost every percentile. Thus, most requests are treated
slower. At the median, the latency increases by 25\%. The latencies in the tail
of the CDF are explained by the inherent and the temporary slowdowns. We re-run
the experiment using FIO and saw similar results.

We also measure latency when sending one request at a time. We vary request
size between 128KB and 1MB. We find that single request latency also increases
after permanent slowdown and for all sizes. The increase in latency is constant
in absolute terms and is thus not correlated to request size. The latency for the
default request size of 256 KB increased by 33\%.  These findings show that
latency increases are due to the drive and not due to the software layers.

\noindent{\bf Discussion on internal SSD root cause.\hspace{0.2in}}
The read disturbance and retention errors discussed as potential culprits for
temporary slowdown could conceivably lead to permanent
slowdown~\cite{mutlu_errors_date12} if left uncorrected by the drive. However,
the same argument we made for temporal slowdown applies. Read disturbance and
retention occur on drives much more worn out (orders of magnitude more P/E
cycles) than ours and after performing orders of magnitude more reads.  We
hypothesize that the reason for permanent slowdown lies with error correction
algorithms being triggered inside the drive after enough time has passed since
file creation.

\subsection{Mitigation Strategies}\label{ssec:permMitigation}

We consider mitigation strategies that are more aggressive in generating
request level parallelism in the hope that they could compensate for the
throughput loss. We find that both direct IO as well as increasing the number
of requests sent in parallel with buffered IO can mask permanent slowdown.

\begin{figure}[t]
	\centering
	\includegraphics[width=\linewidth]{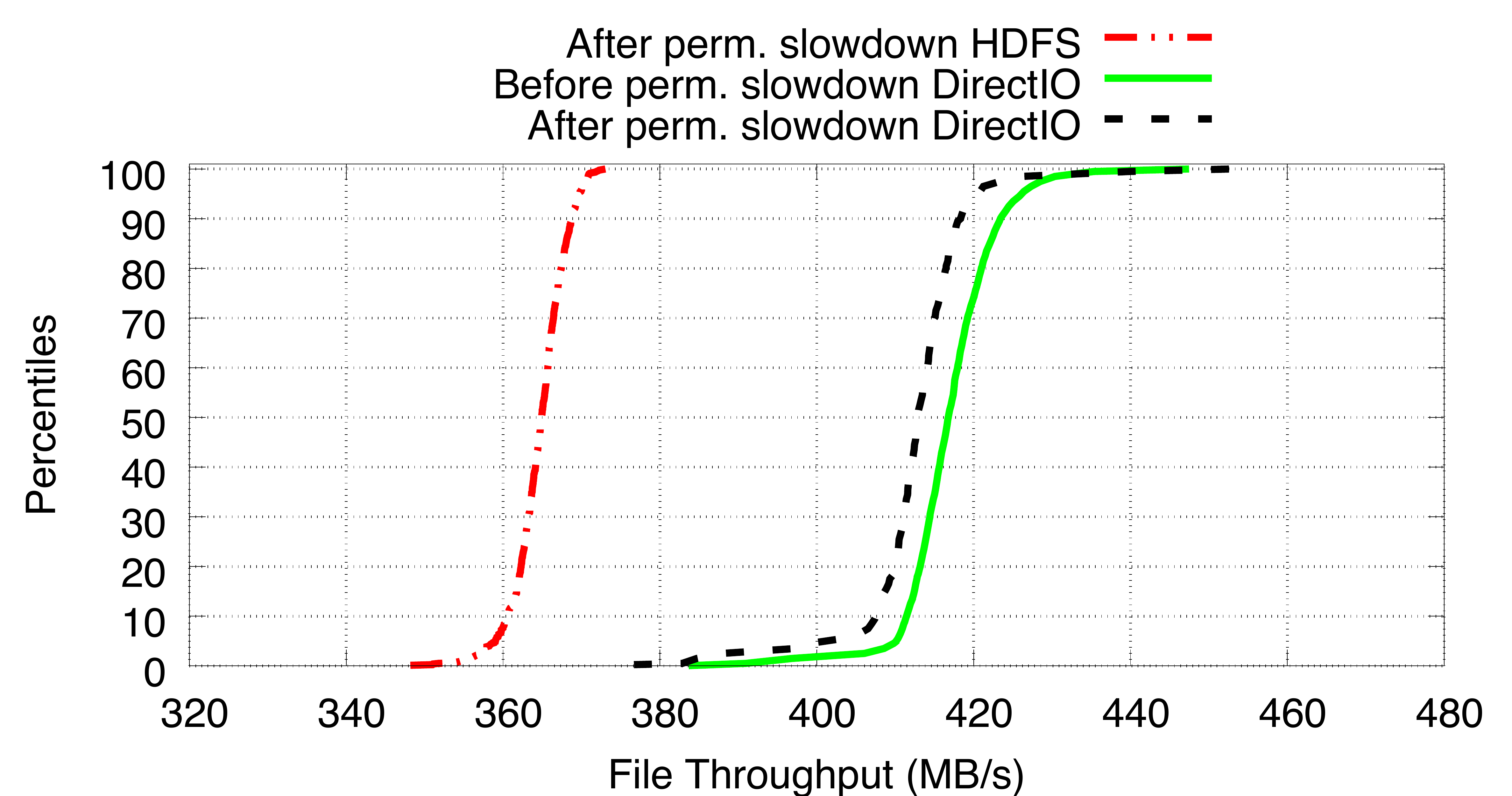}
	\caption{HDFS read vs FIO DirectIO. CDFs of file throughput before and after permanent slowdown.}
	\label{fig:cdf_thpt_hdfs_directIO} 
	\vspace{-.2in}
\end{figure}

First, we look at the behavior of permanent slowdown when reading with direct
IO. It is known that direct IO issues more and larger requests to the block
layer, when compared to buffered IO~\cite{he2017unwritten}. In our experiments
it issues four 1MB in parallel. Figure~\ref{fig:cdf_thpt_hdfs_directIO}
presents a throughput comparison between HDFS and FIO with direct IO.  The two
rightmost CDFs correspond to the throughput of FIO with direct IO before and
after permanent slowdown. The difference between the two is minimal. We
repeated the experiment using a smaller, 256KB direct IO request size (by tuning
\texttt{/sys/block/\textless device\textgreater/queue/max\_sectors\_kb)}.  The results remained the same
suggesting that having a larger number of parallel requests is key for
best performance. 

\begin{figure}[t]
	\centering
	\includegraphics[width=\linewidth]{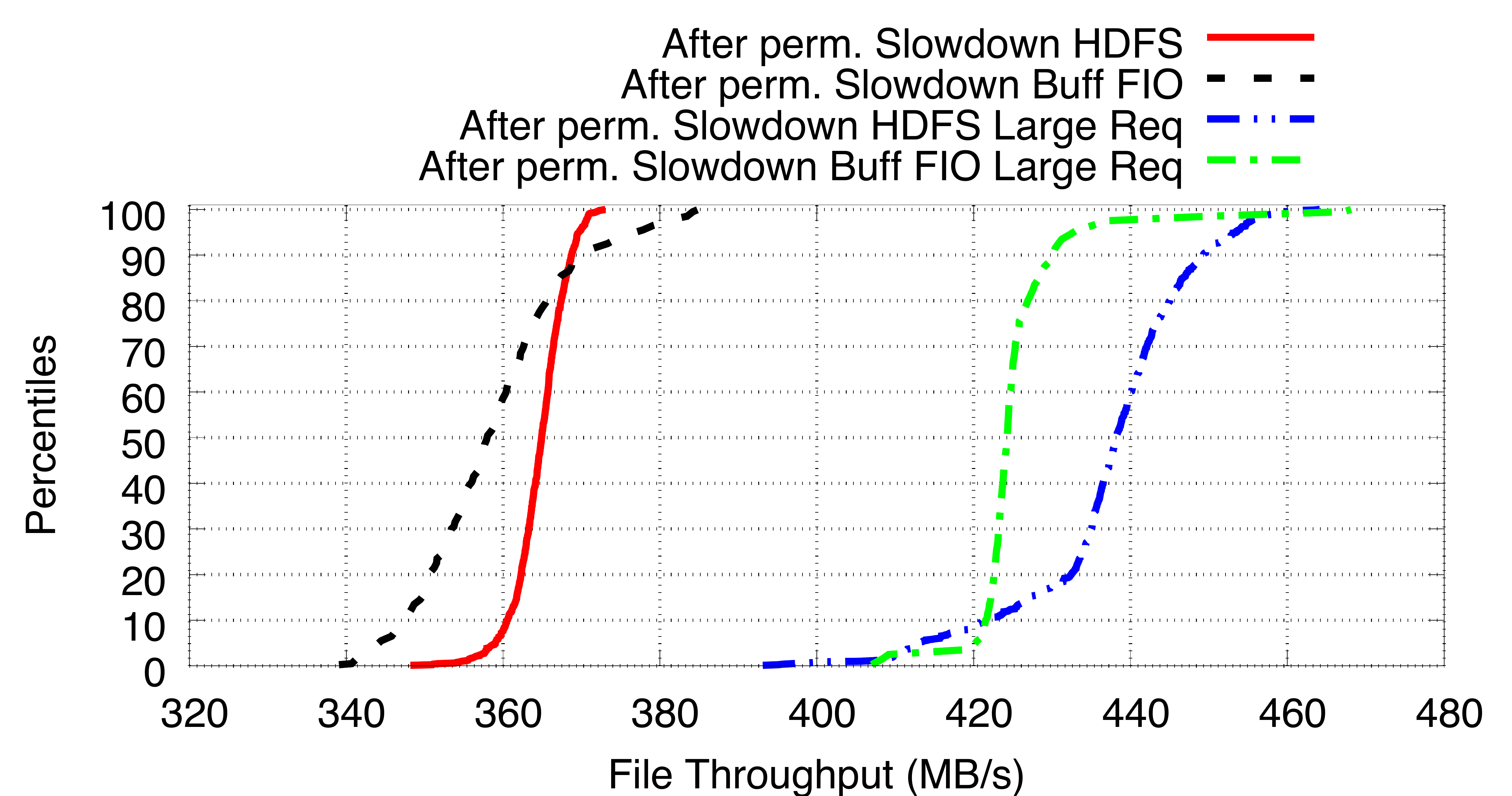}
	\caption{HDFS read vs Buffered FIO. CDFs of file througput. Before permanent slowdown
	with IO requests of 256KB, and after permanent slowdown with large IO requests of 2MB.}
	\label{fig:cdf_thpt_hdfs_fio_large_req} 
	\vspace{-.2in}
\end{figure}

We also analyzed making buffered IO more aggressive.  We increase the request
size from 256KB to 2MB by modifying the read ahead value.  This change
automatically brings about a change in the number of request sent in parallel
to the drive.  When request size is 256KB, two requests execute in parallel. For
a 2MB request, four parallel execute in parallel.
Figure~\ref{fig:cdf_thpt_hdfs_fio_large_req} presents four CDFs representing
the throughput after permanent slowdown with HDFS reads and FIO buffered reads.
The leftmost CDFs correspond to the default request size of 256KB and show the
impact of permanent slowdown.  The rightmost CDFs are for a request size of
2MB. The modified buffered IO is able to mask the permanent slowdown with
increased parallelism.   

\section{Discussion}
\label{sec:discussion}

We showed that intrinsic and permanent slowdowns occur because software cannot
adapt its strategy for extracting the maximum performance and parallelism from
the device in the face of changes in SSD behavior. In the common case, software
can extract the maximum performance and parallelism from the SSD using a set
strategy of generating IO requests. When SSD performance drops due to internal
causes, the same set strategy cannot continue to extract the same performance
level. A more aggressive strategy is needed. Yet, HDFS cannot
adapt. This points to the need to consider more adaptable software designs that
readjust according to perceived performance drops and instabilities in
hardware.

The more aggressive approaches that we evaluated were switching to direct IO
and increasing the size and number of parallel IO requests in buffered IO.
Unfortunately, for existing applications, especially those with large code
bases like HDFS, these more aggressive approach may not always be easy to
leverage. Switching to direct IO may require extensive changes to application
code. Increasing the aggressiveness of buffered IO may lead to wasted disk
reads if turned on at machine level. If turned on per application, aggressive
buffered IO may influence fairness in the co-existence with collocated
workloads. In addition, from an operational perspective, increasing the
aggressiveness of buffered IO is not straight-forward. First, it is not
intuitive because under the common case the default strategy for buffered IO is
enough to extract maximum performance from SSDs. Moreover, in Linux, the
settings required to increase aggressiveness are controlled by a seemingly
unrelated configuration that controls read-ahead size.

Our findings have an impact on they way systems are benchmarked on SSDs.  If
two systems are tested on copies of the exact same file, 10-20\% of the
performance difference may come from intrinsic slowdown (a copy is more
fragmented) and/or permanent slowdown (a copy is older). Even if the same input
file is used but at different points in time, 10\% of the performance
difference may come from permanent slowdown. Finally, if systems are tested for
short periods of time, 10\% of the difference can come from temporary slowdown
if one of the systems is unlucky to run during one slowdown episode. In the
extreme case, one system may be affected by all three slowdowns at the same
time while another may only be slightly affected by intrinsic slowdown. In this
case, almost 30\% of the performance difference may come from the slowdowns and
not the systems under test.

\section{Related Work}
\label{sec:related}

\noindent{\bf Related to sources of performance variation internal to SSDs.\hspace{0.2in}} 
Garbage collection (GC) in SSDs is known to trigger temporary slowdowns but it
is write induced~\cite{skourtis2014flash,yan2017tiny}. Flash on
Rails~\cite{skourtis2014flash} reports no GC-like effects in read-only
workloads. Since our paper focus solely on read-only workloads we do not
discuss further GC.

There are two types of errors that can appear in read-only workloads, retention
errors and read errors.  Retention errors occur when data stored in a cell
changes as time passes and are caused by the charge in a cell dissipating over
time through the leakage current~\cite{mutlu_errors_date12,
mutlu_retention_hpca15}. Read (disturbance) errors occur when the data in a
cell is modified over time as a neighboring cell is read repeatedly and are
caused by the repeated reads shifting the threshold voltages of unread cells
and switching them to a different logical state~\cite{mutlu_errors_date12,
mutlu_disturb_dsn15}. In practice, retention errors happen much more frequently
than read disturbance errors~\cite{mutlu_facebookstudy_sigmetrics15}.

The temporary slowdowns we encountered show a different pattern compared to the
two read errors described above. Related work shows that read errors are
highly correlated with the number of P/E cycles that the drive went
through~\cite{mutlu_errors_date12}. Our drives have a very low P/E cycle.  At
the end of our experiments, the amount of data written to the drives over their
entire lifetime was just 1TB, double their capacity. In contrast, related work
uses drives with thousands of P/E cycles to show a noticeable increase in error
rates~\cite{mutlu_errors_date12}. Similarly, to obtain read errors, related
work~\cite{mutlu_disturb_dsn15} perform hundreds of thousands of reads on a
single page in order to see noticeable effects. Our experiments perform at most
a few thousand reads. In addition, the read-errors results from related
work~\cite{mutlu_disturb_dsn15} are on drives that already underwent thousands
of P/E cycles. 

Gunawi~\textit{et al.}~\cite{gunawi2018fail} study 101 reports of fail-slow hardware
(some of which SSD-related) incidents, collected from large-scale cluster
deployments. One the SSD front, they find firmware bugs that cause latency
spikes or stalls and slow reads due to read retries or parity-based read
reconstruction. The study finds that slow reads occurs mostly on worn out SSDs
or SSDs that approach end of life. We show that similar problems can occur on
very lightly used SSDs. Moreover, we analyze the impact that these hardware
issues have at the application level.

Jung~\textit{et al.}~\cite{revisiting_sigmetrics13} find at least 5x increased
latency on reads when enabling reliability management on reads (RMR). RMR
refers collectively to handling read disturbance management, runtime bad block
management, and ECC. The latency differences causing the slowdown we uncover
are much less pronounced. Moreover, our slowdowns illustrate dynamics in read
latency over time whereas this work focuses on read-related
insights from parameter sweeps.

Hao~\textit{et al.}~\cite{hao2016tail} perform a large-scale study analysis
of tail latency in production HDDs and SSDs.
Their study presents a series of slowdowns and shows that drive internal 
characteristics are most likely responsible form them. They characterize
long slowdown periods that (may) last hours and affect the whole drive,
without any particular correlation to IO rate. 
Like them, we find that the drive is most likely responsible for most of the
slowdowns. We did not experience large period of slowdowns
across the whole drive, probably due to the fact that our drives 
are more lightly used.


\noindent{\bf Related to fragmentation in SSDs.\hspace{0.2in}} 
Conway~\textit{et al.}~\cite{conway2017file} show that certain workloads cause
file systems to age (become fragmented) and this causes performance loss even
on SSDs. Their workloads involve many small files (\textless 1MB).  
Kadekodi~\textit{et al.}~\cite{kadekodi2018geriatrix} also exposes the impact of
aging in SSD across a variation of workloads and file sizes. They
focus on replicating fragmentation to improve benchmarking quality.
Similarly, Chopper~\cite{he2015reducing} studies tail latencies introduced by block
allocation in ext4 in files of maximum 256 KB. In contrast, we study intrinsic
slowdown in much larger files (256 MB) and we quantify the impact on HDFS.

\noindent{\bf Related to extracting best performance out of SSDs.\hspace{0.2in}} 
He~\textit{et al.}~\cite{he2017unwritten} focus on five unwritten rules that
applications should abide by to get the most performance out of the SSDs and
analyze how a number of popular applications abide by those rules.  These rules
boil down to specific ways of creating and writing files: write aligned, group
writes by death time, create data with similar lifetimes, etc. These findings
are all complementary to our work. The authors also point to small IO request
sizes and argue that they are unlikely to use the SSD parallelism well.  In
contrast we see that in the common case, the default IO request sizes can
extract the maximum SSD performance but fall short when hardware behavior
changes as exemplified by the permanent slowdown.

\noindent{\bf Related to storage-influenced HDFS performance\hspace{0.2in}} 
Shafer~\textit{et al.}~\cite{shafer2010hadoop} analyze the performance of HDFS
v1 on HDDs using Hadoop jobs. They show three main findings. First,
architectural bottlenecks exist in Hadoop that result in inefficient HDFS
usage. Second, portability limitations prevent Java from exploiting
features of the native platform. Third, HDFS makes assumptions about how native
platforms manage storage resources even though these vary widely in design and
behavior. Our findings complement this past work by looking at SSDs
instead of HDDs. Moreover, we look at the influence that internal drive
characteristics have on HDFS performance while this past work focuses on
software-level interactions.
Harter~\textit{et al.}~\cite{harter2014analysis} study HDFS behavior under
HBase workload constraints: store small files (\textless 15MB) and random IO.
In contrast, we study HDFS under regular conditions, with large files and
sequential IO. 

\noindent{\bf Related to performance variability of storage stacks\hspace{0.2in}} 
Cao~\textit{et al.}~\cite{cao2017performance} study the performance variation
of modern storage stacks, on both SSDs and HDDs. For the workloads they
analized they find ext4-SSD performance to be stable even across different
configurations, with less than 5\% relative range. In contrast we show
variations of up to 30\% over time, for one single configuration for HDFS.
Maricq~\textit{et al.}~\cite{maricq2018taming} conduct a large-scale 
variability study. Storage wise, they focus on understanding 
performance variability between HDDs and SSDs. Similar to us, they find
that sending large number of requests to the SSDs reduces performance variability.
However, they focus on workloads with direct IO and small request sizes (4KB). 
In contrast, we study SSD variability both under direct IO and buffered IO. We
dive deeper into the importance on the number and size of requests.
Vangoor~\textit{et al.}~\cite{vangoor2017fuse} analyze the performance overheads
of FUSE versus native ext4. Their analysis shows that in some cases FUSE overhead
is negligible, while in some others it can heavily degrade performance. HDFS is
also a user space file system, however it has a different architecture and
functionality, and use cases than FUSE. In this work, we analyze the interaction 
between HDFS and lower layers of the storage stack, 
under HDFS main use case, sequential IO in large files.

\section{Conclusion}
\label{sec:conclusion}

In this paper we introduced and analyzed three surprising performance problems
(inherent, temporal and permanent slowdowns) that stop HDFS from extracting
maximum performance from some SSDs. These problems are introduced by the layers
sitting beneath HDFS (file system, SSDs). The lower layers also hold the key to
masking two of the three problems by increasing IO request parallelism during
the problems. Unfortunately, HDFS does not have the ability to adapt. Its
access pattern successfully extracts maximum performance from SSDs in the
common case but it is not aggressive enough to mask the performance problems we
found. Our results point to a need for adaptability in storage stacks.

{\normalsize \bibliographystyle{acm}
\bibliography{usenix2019}}

\begin{thebibliography}{10}

\bibitem{nutanix-platforms}
{Nutanix Hardware Platforms}.
\newblock \url{https://www.nutanix.com/products/hardware-platforms/}.

\bibitem{bonwick2003zettabyte}
{\sc Bonwick, J., Ahrens, M., Henson, V., Maybee, M., and Shellenbaum, M.}
\newblock The zettabyte file system.
\newblock In {\em Proc. of the 2nd Usenix Conference on File and Storage
  Technologies\/} (2003), vol.~215.

\bibitem{mutlu_errors_date12}
{\sc Cai, Y., Haratsch, E.~F., Mutlu, O., and Mai, K.}
\newblock Error patterns in mlc nand flash memory: Measurement,
  characterization, and analysis.
\newblock In {\em Proceedings of the Conference on Design, Automation and Test
  in Europe DATE 12}.

\bibitem{mutlu_disturb_dsn15}
{\sc Cai, Y., Luo, Y., Ghose, S., and Mutlu, O.}
\newblock Read disturb errors in mlc nand flash memory: Characterization,
  mitigation, and recovery.
\newblock In {\em Proceedings of the 2015 45th Annual IEEE/IFIP International
  Conference on Dependable Systems and Networks DSN 15}.

\bibitem{mutlu_retention_hpca15}
{\sc Cai, Y., Luo, Y., Haratsch, E.~F., Mai, K., and Mutlu, O.}
\newblock Data retention in {MLC} {NAND} flash memory: Characterization,
  optimization, and recovery.
\newblock In {\em 21st {IEEE} International Symposium on High Performance
  Computer Architecture, {HPCA} 2015, Burlingame, CA, USA, February 7-11,
  2015}.

\bibitem{cao2017performance}
{\sc Cao, Z., Tarasov, V., Raman, H.~P., Hildebrand, D., and Zadok, E.}
\newblock On the performance variation in modern storage stacks.
\newblock In {\em FAST\/} (2017), pp.~329--344.

\bibitem{conway2017file}
{\sc Conway, A., Bakshi, A., Jiao, Y., Jannen, W., Zhan, Y., Yuan, J., Bender,
  M.~A., Johnson, R., Kuszmaul, B.~C., Porter, D.~E., et~al.}
\newblock File systems fated for senescence? nonsense, says science!
\newblock In {\em FAST\/} (2017), pp.~45--58.

\bibitem{tail_at_scale}
{\sc Dean, J., and Barroso, L.~A.}
\newblock The tail at scale.
\newblock {\em Commun. ACM 56}, 2 (Feb. 2013), 74--80.

\bibitem{gfs}
{\sc Ghemawat, S., Gobioff, H., and Leung, S.-T.}
\newblock The google file system.
\newblock In {\em Proceedings of the Nineteenth ACM Symposium on Operating
  Systems Principles}, SOSP '03.

\bibitem{gunawi2018fail}
{\sc Gunawi, H.~S., Suminto, R.~O., Sears, R., Golliher, C., Sundararaman, S.,
  Lin, X., Emami, T., Sheng, W., Bidokhti, N., McCaffrey, C., et~al.}
\newblock Fail-slow at scale: Evidence of hardware performance faults in large
  production systems.
\newblock {\em ACM Transactions on Storage (TOS) 14}, 3 (2018), 23.

\bibitem{hao2016tail}
{\sc Hao, M., Soundararajan, G., Kenchammana-Hosekote, D.~R., Chien, A.~A., and
  Gunawi, H.~S.}
\newblock The tail at store: A revelation from millions of hours of disk and
  ssd deployments.
\newblock In {\em FAST\/} (2016), pp.~263--276.

\bibitem{harter2014analysis}
{\sc Harter, T., Borthakur, D., Dong, S., Aiyer, A.~S., Tang, L.,
  Arpaci-Dusseau, A.~C., and Arpaci-Dusseau, R.~H.}
\newblock Analysis of hdfs under hbase: a facebook messages case study.
\newblock In {\em FAST\/} (2014), vol.~14, p.~12th.

\bibitem{he2017unwritten}
{\sc He, J., Kannan, S., Arpaci-Dusseau, A.~C., and Arpaci-Dusseau, R.~H.}
\newblock The unwritten contract of solid state drives.
\newblock In {\em Proceedings of the Twelfth European Conference on Computer
  Systems\/} (2017), ACM, pp.~127--144.

\bibitem{he2015reducing}
{\sc He, J., Nguyen, D., Arpaci-Dusseau, A.~C., and Arpaci-Dusseau, R.~H.}
\newblock Reducing file system tail latencies with chopper.
\newblock In {\em FAST\/} (2015), vol.~15, pp.~119--133.

\bibitem{revisiting_sigmetrics13}
{\sc Jung, M., and Kandemir, M.}
\newblock Revisiting widely held ssd expectations and rethinking system-level
  implications.
\newblock In {\em Proceedings of the ACM SIGMETRICS/International Conference on
  Measurement and Modeling of Computer Systems SIGMETRICS 13}.

\bibitem{kadekodi2018geriatrix}
{\sc Kadekodi, S., Nagarajan, V., Ganger, G.~R., and Gibson, G.~A.}
\newblock Geriatrix: Aging what you see and what you don’t see. a file system
  aging approach for modern storage systems.
\newblock In {\em Proceedings of the 2018 USENIX Conference on Usenix Annual
  Technical Conference\/} (2018), USENIX Association, pp.~691--703.

\bibitem{lee2015f2fs}
{\sc Lee, C., Sim, D., Hwang, J.~Y., and Cho, S.}
\newblock F2fs: A new file system for flash storage.
\newblock In {\em FAST\/} (2015), pp.~273--286.

\bibitem{maricq2018taming}
{\sc Maricq, A., Duplyakin, D., Jimenez, I., Maltzahn, C., Stutsman, R., and
  Ricci, R.}
\newblock Taming performance variability.
\newblock In {\em 13th $\{$USENIX$\}$ Symposium on Operating Systems Design and
  Implementation ($\{$OSDI$\}$ 18)\/} (2018), pp.~409--425.

\bibitem{mathur2007new}
{\sc Mathur, A., Cao, M., Bhattacharya, S., Dilger, A., Tomas, A., and Vivier,
  L.}
\newblock The new ext4 filesystem: current status and future plans.
\newblock In {\em Proceedings of the Linux symposium\/} (2007), vol.~2,
  pp.~21--33.

\bibitem{mutlu_facebookstudy_sigmetrics15}
{\sc Meza, J., Wu, Q., Kumar, S., and Mutlu, O.}
\newblock A large-scale study of flash memory failures in the field.
\newblock In {\em Proceedings of the 2015 ACM SIGMETRICS International
  Conference on Measurement and Modeling of Computer Systems}.

\bibitem{mytkowicz2009producing}
{\sc Mytkowicz, T., Diwan, A., Hauswirth, M., and Sweeney, P.~F.}
\newblock Producing wrong data without doing anything obviously wrong!
\newblock {\em ACM Sigplan Notices 44}, 3 (2009), 265--276.

\bibitem{ousterhout2018always}
{\sc Ousterhout, J.}
\newblock Always measure one level deeper.
\newblock {\em Commun. ACM 61}, 7 (June 2018), 74--83.

\bibitem{rodeh2013btrfs}
{\sc Rodeh, O., Bacik, J., and Mason, C.}
\newblock Btrfs: The linux b-tree filesystem.
\newblock {\em ACM Transactions on Storage (TOS) 9}, 3 (2013), 9.

\bibitem{shafer2010hadoop}
{\sc Shafer, J., Rixner, S., and Cox, A.~L.}
\newblock The hadoop distributed filesystem: Balancing portability and
  performance.
\newblock In {\em Performance Analysis of Systems \& Software (ISPASS), 2010
  IEEE International Symposium on\/} (2010), IEEE, pp.~122--133.

\bibitem{hdfs-paper}
{\sc Shvachko, K., Kuang, H., Radia, S., and Chansler, R.}
\newblock {The Hadoop Distributed File System}.
\newblock In {\em MSST 2010}.

\bibitem{sites2018benchmarking}
{\sc Sites, R.~L.}
\newblock Benchmarking" hello, world!
\newblock {\em Benchmarking 16}, 5 (2018).

\bibitem{skourtis2014flash}
{\sc Skourtis, D., Achlioptas, D., Watkins, N., Maltzahn, C., and Brandt,
  S.~A.}
\newblock Flash on rails: Consistent flash performance through redundancy.
\newblock In {\em USENIX Annual Technical Conference\/} (2014), pp.~463--474.

\bibitem{sweeney1996scalability}
{\sc Sweeney, A., Doucette, D., Hu, W., Anderson, C., Nishimoto, M., and Peck,
  G.}
\newblock Scalability in the xfs file system.
\newblock In {\em USENIX Annual Technical Conference\/} (1996), vol.~15.

\bibitem{vangoor2017fuse}
{\sc Vangoor, B. K.~R., Tarasov, V., and Zadok, E.}
\newblock To fuse or not to fuse: Performance of user-space file systems.
\newblock In {\em FAST\/} (2017), pp.~59--72.

\bibitem{yan2017tiny}
{\sc Yan, S., Li, H., Hao, M., Tong, M.~H., Sundararaman, S., Chien, A.~A., and
  Gunawi, H.~S.}
\newblock Tiny-tail flash: Near-perfect elimination of garbage collection tail
  latencies in nand ssds.
\newblock {\em ACM Transactions on Storage (TOS) 13}, 3 (2017), 22.

\bibitem{rdd}
{\sc Zaharia, M., et~al.}
\newblock Resilient distributed datasets: A fault-tolerant abstraction for
  in-memory cluster computing.
\newblock In {\em NSDI 2012}.

\end{thebibliography}


\end{document}